\documentclass[a4paper,12pt]{scrartcl}

\pdfoutput=1

\usepackage{lineno}
\usepackage[utf8]{inputenc}
\usepackage[T1]{fontenc}

\usepackage{amsmath,amsthm,amsopn,xspace,bm}
\usepackage{array}

\usepackage{cite}

\usepackage[english]{babel}

\usepackage{graphicx}

\usepackage{gnuplot-lua-tikz}

\usepackage{skmath}

\usepackage[babel]{microtype}

\usepackage{bera}
\usepackage[charter]{mathdesign}
\newcommand{\VEC}{\bm}
\newcommand{\MAT}{\bm}

\DeclareMathOperator{\sign}{sgn}

\usepackage[colorinlistoftodos,shadow]{todonotes}

\begin{document}

\subject{Preprint}

\title{Roe-type schemes for shallow water magnetohydrodynamics with
  hyperbolic divergence cleaning}

\author{Friedemann Kemm\footnote{Brandenburg University of Technology,
    Platz der Deutschein Einheit 1, 03046 Cottbus, Germany,
    kemm@math.tu-cottbus.de}}

\maketitle

\begin{abstract}
  We discuss Roe-type linearizations for non-conservative shallow
  water magnetohydrodynamics without and with hyperbolic divergence
  cleaning.
\end{abstract}

\section{Introduction}
\label{sec:introduction}

Hyperbolic conservation laws are usually equipped with additional
conditions. Most important is the existence of a convex entropy---in
the case of shallow water the energy---which singles out the
physically relevant solution from of the large set of possible weak
solutions. Sometimes, especially when there is no convex entropy, or
the system degenerates into a weakly or resonant hyperbolic system
like shallow water magnetohydrodynamics (SMHD), other laws have to be
included to find physical solutions. In the first case (convex
entropy), the additional law is introduced to accommodate an
additional variable, namely the entropy, which depends on the state
variables, but is no state variable itself like the energy in shallow
water flows. In the latter case, we have additional partial
differential equations for the state variables themselves. In the
first case, the additional law is a partial differential equation or
inequality of evolution type, usually a conservation law itself, in
the latter, it is a first order non-evolutionary constraint. These
additional constraints are, as Dafermos points
out~\cite{dafermosbuch,dafermospaper}, involutions for the underlying
system of conservation laws. So the resulting system, which includes
both the evolution system and the condition, has more equations than
unknowns. If the involution is satisfied by the initial state for the
evolution equations, it is satisfied by the solution of the evolution
system for all times. Thus, in the continuous setting, the constraint
is merely a condition for the initial state. The equations
of shallow water magnetohydrodynamics (SMHD) are of this latter
type, which is a property inherited from full MHD.

To cope with such constraints, there are different approaches in the
literature. First, there are approaches, which are designed to model
the constraint numerically. Many of them are done on staggered
grids~\cite{balsdiv,spicer,hawley88}. Some newer approaches also work
on collocated
grids~\cite{manfey,Manuel,rossmanith06,mishra-I,mishra-II,mishra-III,fuchs-induction,fuchs-split,waagan}
or in the context of Discontinuous-Galerkin
schemes~\cite{besse05}. Usually this class of schemes is referred to
as \emph{constrained transport} (CT). Applications to SMHD were
provided, e.\,g., by de~Sterck~\cite{hans-constrained} and
Rossmanith~\cite{rossmanith}. The basic strategy for most of these
CT-methods is to perform a time step for the conservative evolution
equations (standard model). In a second step, the magnetic field is
recomputed on the basis of a different equation, quite often on the
evolution equation for the vector potential of the magnetic field. As
a consequence, the evolution of the magnetic field is, at least
formally, non-conservative. An alternative approach is the one
provided by Fey and Torrilhon~\cite{manfey,Manuel}. It is based on
averaging fluxes of neighbouring Riemann problems, a strategy that is
well known to heavily increase numerical viscosity. With the other
CT-methods, it shares the drawback that a change in the underlying
scheme---e.\,g.\ increasing the order of accuracy---also requires a
new development of the constraint preserving strategy. Therefore, we
concentrate on a different approach.

A second family of schemes is based on a modification of the system of
partial differential equations which makes up the constraint part of
the evolution system itself. In the context of plasma physics, a
popular approach is to transport the involution term, in this case the
divergence of the magnetic field, with the flow velocity. This was
first suggested for numerical simulations of full MHD by Brackbill and
Barnes~\cite{brackbill1} and put forward by employing Godunovs full
symmetrizable form\footnote{It is interesting to note that this form
  was first discovered by Godunov~\cite{godunov1972symmetric} as
  symmetrizable form of MHD and then rediscovered by Powell
  et~al.~\cite{powell95} as Galilean invariant form of MHD. Since it
  has an entropy~\cite{godunov1972symmetric,dafermosbuch}, one would
  not need any involution for the system.} of the
MHD-equations~\cite{godunov1972symmetric} by Powell
et~al.~\cite{powell95,linde99}. In~\cite{fuchs-induction,fuchs-split,waagan},
this approach is even combined with constrained transport. Another
possibility is to apply a kind of a Generalized Lagrange Multiplier (GLM)
approach~\cite{eric}, a method which can appear in several variants:
a Hodge-projection scheme,  a parabolic
treatment of the involution term as was suggested by
Marder~\cite{Marder87},  a hyperbolic system\,--\,the
involution term is radiated with an artificial speed out of the
computational domain~\cite{timedomain,maxjcp99}\,--\,or it results in
a treatment of the involution in the manner of a telegraph
equation~\cite{divpaper,porquerolles} (Crockett et~al.~\cite{Crockett}
even combine the Marder approach with a Hodge-projection method.) In
the context of electromagnetic models and plasma physics these
approaches are usually referred to as \emph{divergence
  cleaning}. Although possible in principle, the combination of the
transport of the involution with the hyperbolic GLM approach is, to
our knowledge, not yet described in the literature\footnote{But it seems to be
in practical use, as we can conclude from several oral communications
in the last years.}. 

It is worth noting that, as Brackbill and Barnes~\cite{brackbill1}
point out, the standard system of MHD is deduced from physical
principles together with the assumption that the magnetic field is
solenoidal. Without this a~priori assumption, one would simply end up
with the Powell system. As we pointed out in our earlier
work~\cite{porquerolles,pasadena,anume}, the derivation of the Maxwell
equations themselves without the neglect of magnetic charge and
magnetic current, just on the basis of physical symmetries, would lead
to the standard Maxwell equations augmented by the terms of the
hyperbolic GLM correction. Thus, from the physical point of view, it
is desirable to employ MHD (and SMHD) equations which are based on the
combination of Powell and hyperbolic GLM.

In this paper, we develop (and test) Roe-type schemes for hyperbolic
systems of SMHD equations. For the conservative standard equations, as
introduced by Gilman~\cite{gilman-smhd}, the characteristic analysis
was performed by de~Sterck~\cite{hans1} and a Roe-type scheme
developed by Rossmanith~\cite{rossmanith}. Based on this preparatory
work, we investigate the hyperbolic structure of the extended SMHD
systems, i.\,e.\, with Powell correction and/or hyperbolic GLM
correction, in the following section. In
Section~\ref{sec:roe-matrices-smhd} we discuss possibilities to get
Roe matrices for these systems. This again will be based on the work
by Rossmanith~\cite{rossmanith}. In
Section~\ref{sec:roe-matrices-smhd} the resulting schemes are tested
at hand of two prototypical test cases and compared to the scheme for
the standard equations. Furthermore, we also cover the choice for
the artificial wave speed and the numerical viscosity of the waves
which radiate the divergence errors.

\section{The governing equations, combining Powell with GLM-correction}
\label{sec:govern-equat-comb}

Here, we discuss the shallow water MHD (SMHD) equations
with and without Powell and GLM-correction and their eigensystems.  
The eigensystems are needed in later sections to set up the Roe-type
schemes for the different SMHD models.

\subsection{Conservative equations}
\label{sec:equat-with-glm}

Before we introduce the GLM-correction, we discuss the standard
equations and their modification according to Powell. The latter has
the desirable property to be fully hyperbolic and not only resonant
hyperbolic.

\subsubsection{Without divergence cleaning}
\label{sec:standard-equations}

In~\cite{gilman-smhd}, Gilman argues that the classical ``shallow
water'' equations of geophysical fluid dynamics should be useful for
studying the global dynamics of the solar tachocline and demonstrates
the existence of an MHD analogon that would allow taking into account
the strong toroidal magnetic field likely to be present there. So he
presents a derivation analogous to that for the classical shallow
water equations and comes up with the following system of shallow
water magnetohydrodynamics (SMHD) equations
\begin{align}
  \label{eq-smhd:65}
    h_t + \nabla \cdot [ h \VEC v] & = 0\;, \\
  (h \VEC v)_t + \nabla \cdot [ h \VEC{v\circ v} - h
  \VEC{B\circ B} +
  \frac{g h^2}{2} \MAT I ] & = \VEC 0\;, \label{eq:1} \\
  (h \VEC B)_t - \nabla \times [ \VEC v \times
  (h \VEC B)] &  = \VEC 0\;, \label{eq:2} \\ 
  \label{eq-smhd:66}
  \nabla \cdot (h \VEC B) & = 0\;,
\end{align}
which inherits most of its behaviour from the original MHD-system. The
main difference is that, due to the averaging over the third space
dimension, the magnetic field \(\VEC B\) is now replaced by \(h \VEC
B\), where \(h\) denotes the height of the present fluid layer, and
\(g\) is the gravitational constant. Again, the last
equation~\eqref{eq-smhd:66} is a constraint for the evolution
system~\eqref{eq-smhd:65}~--~\eqref{eq:2} and does not contribute to
the evolution of the solution. For the hyperbolic structure, we only
consider the evolution system~\eqref{eq-smhd:65}~--~\eqref{eq:2}.

Because the system is rotationally invariant, we therefore consider in
the following the flux Jacobian only for the first space-direction,
which reads as
\begin{equation}
  \label{eq:3}
  \MAT A_\text{standard} =
  \begin{pmatrix}
    0 & 1 & 0 & 0 & 0 \\
    -u^2 + B_1^2 + gh & 2u & 0 & -2B_1 & 0 \\
    -uv + B_1 B_2 & v & u & -B_2 & -B_1 \\
    0 & 0 & 0 & 0 & 0 \\
    vB_1 - uB_2 & B_2 & -B_1 & -v & u
  \end{pmatrix}\;. 
\end{equation}


The wave speeds, i.\,e.\ the eigenvalues of the flux Jacobian in the
first space direction, are
\begin{equation}
  \label{eq:148}
  \lambda_1 = u-c_{g}\;,\quad \lambda_2 = u-\abs{B_1}\;,\quad
  \lambda_3 = 0\;,\quad \lambda_4 = u+\abs{B_1}\;,\quad \lambda_5 =
  u+c_{g} 
\end{equation}
with the magneto-gravitational speed
\begin{equation}
  \label{eq:149}
  c_{g} = \sqrt{B_1^2 + gh}\;,
\end{equation}
which, for a vanishing magnetic field, would coincide with the
celerity~\(c_g\) of the shallow water equations. 
The third eigenvalue~\(\lambda_3 = 0\) indicates that the system is
not Galilean invariant. For the other wave speeds, 
\begin{equation}
  \label{eq:150}
  \lambda_1 \leq \lambda_2 \leq \lambda_4 \leq \lambda_5
\end{equation}
is satisfied for any state~\(\VEC q\). If the first component of the
magnetic field vanishes,~\(\lambda_2\) and~\(\lambda_4\)
coincide. Thus, the system is not strictly hyperbolic. To decide if
the system is hyperbolic or weakly/resonant hyperbolic, one has to
consider the eigenvectors. The right eigenvectors are\footnote{Note
  that the eigensystem given in this section deviates from the system
  given by de~Sterck~\cite{hans-constrained,hans1}. The reason is that
  de~Sterck investigates the system with the so called \emph{Powell
    correction} which will be discussed in
  Section~\ref{sec:powell-correction}.}
\begin{align}
  \label{eq:151}
  \VEC r_1 & =
  \begin{pmatrix}
    1 \\ u-c_{g} \\ v \\ 0 \\ B_2
  \end{pmatrix}\;,& 
  \VEC r_5 & =
  \begin{pmatrix}
    1 \\ u+c_{g} \\ v \\ 0 \\ B_2
  \end{pmatrix} \\ 
  \intertext{for the magneto-gravitational waves,}
  \VEC r_2 & =
  \begin{pmatrix}
    0 \\ 0 \\ 1 \\ 0 \\ \sign (B_1)
  \end{pmatrix}\;,& 
  \VEC r_4 & =
  \begin{pmatrix}
    0 \\ 0 \\ 1 \\ 0 \\ - \sign (B_1)
  \end{pmatrix} \\ 
\end{align}
for the Alfven waves, and
\begin{equation}
  \label{eq:152}
    \VEC r_3  =
  \begin{pmatrix}
    2 B_1\,(u-B_1)\,(u+B_1) \\ 
    (vB_1-uB_2)\,(u-B_1)\,(u+B_1)+ gh\,(vB_1+uB_2) \\
    -(u-c_{g})\,(u+c_{g})\,(u-B_1)\,(u+B_1) \\
    (B_1B_2 - uv)\,(u-B_1)\,(u+B_1) + gh\,(B_1B_2 + uv)
  \end{pmatrix}\;. 
\end{equation}
for the remaining wave. If for any~\(k\neq 3\) the
eigenvalue~\(\lambda_k\) vanishes, the system is only resonant
hyperbolic. 

The according left eigenvalues are
\begin{equation}
  \label{eq:153}
  \begin{split}
    \VEC l_1 & = \frac{1}{2c_{g}}\, \Bigl(u+c_{g}\,,\, -1\,,\, 0\,,\,
    \frac{2B_1}{u-c_{g}}\,,\, 0 \Bigr)      \\
    \VEC l_2 & = \frac{1}{2}\,\Bigl( -\sign(B_1)B_2 - v\,,\, 0\,,\, 1\,,\,
    \frac{-\sign (B_1)v - B_2}{u-\abs{B_1}}\,,\, \sign (B_1)  \Bigr)  \\
    \VEC l_3 & = \Bigl( 0\,,\, 0\,,\, 0\,,\,
    -\,\frac{1}{(u-c_{g})\,(u+c_{g})\,(u-B_1)\,(u+B_1)}\,,\, 0 \Bigr)   \\
    \VEC l_4 & = \frac{1}{2}\,\Bigl( \sign(B_1)B_2 - v\,,\, 0\,,\, 1\,,\,
    \frac{\sign (B_1)v - B_2}{u-\abs{B_1}}\,,\, -\sign (B_1)   \Bigr)  \\
    \VEC l_5 & = -\,\frac{1}{2c_{g}}\, \Bigl(u-c_{g}\,,\, -1\,,\, 0\,,\,
    \frac{2B_1}{u+c_{g}}\,,\, 0 \Bigr)   
  \end{split}
\end{equation}
These are used in the implementation of the numerical schemes in
clawpack~\cite{clawpack} to perform the numerical experiments for
shallow water MHD in Section~\ref{sec:numerical-tests}. There, we also
employ solvers which are based on the assumption of one-dimensional
physics for the flux normal to a cell face. In 1d,
equation~\eqref{eq-smhd:66} would require~\(hB_1\) to be
constant. Together with the 1d-evolution equation, this would mean
that~\(hB_1\) is just a constant parameter. Thus, one equation can be
dropped from the system~\eqref{eq-smhd:65}~--~\eqref{eq:2}. As a
consequence, the third characteristic field, \(\lambda_3\) together
with its eigenvectors, no longer exists. For the other eigenvectors,
the third component should be omitted. As we demonstrated
in~\cite{orig-div}, the application of solvers based on 1d~physics in a
2d~situation leads to the failure of the scheme due to excessive
divergence errors. 




\subsubsection{With GLM-correction}
\label{sec:with-glm-correction}

If we introduce the GLM-correction as it is described
in~\cite{divpaper} for MHD and adopted to SMHD
in~\cite{anume,pasadena}, we have to add the gradient of~\(h\psi\) to
the evolution of the magnetic field and convert the divergence
constraint (cf.\ equation~\eqref{eq-smhd:66}) into the time evolution
for~\(h\psi\) (cf.\ equation~\eqref{eq:48}, where~\(\psi\) is an
artificial quantity. Formally, it is the potential of the magnetic
current. In physics, the magnetic field is solenoidal. Thus, there is
no magnetic current, and~\(\psi\) is a mere constant. The equations
then read as
\begin{align}
  \label{eq:33}
    h_t + \nabla \cdot [ h \VEC v] & = 0\;, \\
  (h \VEC v)_t + \nabla \cdot [ h \VEC{v\circ v} - h
  \VEC{B\circ B} +
  \frac{g h^2}{2} \MAT I ] & = \VEC 0\;, \label{eq-double:1} \\
  (h \VEC B)_t - \nabla \times [ \VEC v \times
  (h \VEC B)] + \nabla(h\psi) &  = \VEC 0\;, \label{eq-double:2} \\ 
  \label{eq:48}
  (h\psi)_t + c_\psi^2\,\nabla \cdot (h \VEC B) & = 0\;,
\end{align}
where~\(c_\psi\) is a constant artificial wave speed. The goal is to
transport the (nonphysical) magnetic monopoles with that speed out of
the computational domain. Again, we consider the flux Jacobian for the
first space direction:
\begin{equation}
  \label{eq:35}
  \MAT A_\text{GLM} =
  \begin{pmatrix}
        0 & 1 & 0 & 0 & 0 & 0 \\
    -u^2 + B_1^2 + gh & 2u & 0 & -2B_1 & 0 & 0 \\
    -uv + B_1 B_2 & v & u & -B_2 & -B_1 & 0 \\
    0 & 0 & 0 & 0 & 0 & 1 \\
    vB_1 - uB_2 & B_2 & -B_1 & -v & u & 0\\
    0 & 0 & 0 & c^2 & 0 & 0
  \end{pmatrix}\;. 
\end{equation}
Compared to~\(\MAT A_\text{standard}\) (cf.\ equation~\eqref{eq:3}), a
sixth row and column was added. However, the eigensystem is changed
considerably. The eigenvectors are eigenvalues to
\begin{equation}
  \label{eq:36}
  \lambda_1 = -c\;,\quad \lambda_2\;, = u-c_g\;, \quad \lambda_3 =
  u-\abs{B_1}\;,\quad \lambda_4 = u+\abs{B_1}\;,\quad \lambda_5\;, =
  u+c_g\;, \quad  \lambda_6 = c\;,
\end{equation}
just as expected. But the eigenvectors are more
complicated. The right eigenvectors for the Alfven waves are only
augmented by a vanishing sixth component and renumbered according
to~\eqref{eq:36}. The same is true for the magnetogravitational
waves. The right eigenvectors for the waves transporting the
divergence errors outside of the computational domain are
\setlength{\extrarowheight}{.8em}
\begin{equation}
  \label{eq:37}
  \VEC r_1 =
  \begin{pmatrix}
    -\,\frac{2 B_1}{c_\psi\left(c_g^2 - (u + c_\psi)^2\right)} \\
    \frac{2 B_1}{c_g^2 - (u + c_\psi)^2} \\
    - \, \frac{(u + c_\psi)\left[B_2\left((u + c_\psi)^2 + c_g^2\right) -
      B_1 v (u + c_\psi)\right] + v B_1 (2B_1^2 - c_g^2)}{c_\psi
      \left(B_1^2 - (u + c_\psi)^2\right)\left(c_g^2 - (u + c_\psi)^2\right) }\\
    -\,\frac{1}{c} \\
    - \, \frac{(u + c_\psi)\left[v\left((u + c_\psi)^2 + c_g^2\right) -
      B_1 B_2 (u + c_\psi)\right] + B_1 B_2 (2B_1^2 - c_g^2)}{c_\psi
      \left(B_1^2 - (u + c_\psi)^2\right)\left(c_g^2 - (u + c_\psi)^2\right) }\\
    1
  \end{pmatrix} \\ 
\end{equation}
and
\begin{equation}
  \VEC r_6  =
  \begin{pmatrix}
    \frac{2 B_1}{c_\psi\left(c_g^2 - (u - c_\psi)^2\right)} \\
    \frac{2 B_1}{c_g^2 - (u - c_\psi)^2} \\
    \frac{(u - c_\psi)\left[B_2\left((u - c_\psi)^2 + c_g^2\right) -
      B_1 v (u - c_\psi)\right] + v B_1 (2B_1^2 - c_g^2)}{c_\psi
      \left(B_1^2 - (u - c_\psi)^2\right)\left(c_g^2 - (u -
        c_\psi)^2\right) }\\ 
    \frac{1}{c} \\
    \frac{(u - c_\psi)\left[v\left((u - c_\psi)^2 + c_g^2\right) -
      B_1 B_2 (u - c_\psi)\right] + B_1 B_2 (2B_1^2 - c_g^2)}{c_\psi
      \left(B_1^2 - (u - c_\psi)^2\right)\left(c_g^2 - (u -
        c_\psi)^2\right) }\\ 
    1
  \end{pmatrix}\;.
\end{equation}
On the other hand, the corresponding left eigenvectors for the first
and sixth wave are rather simple:
\begin{equation}
  \label{eq:38}
  \begin{split}
    \VEC l_1 & = \frac{1}{2}\,(0,0,0,-c_\psi,o,1)^T\\
    \VEC l_6 & = \frac{1}{2}\,(0,0,0,c_\psi,o,1)^T\;.
  \end{split}
\end{equation}
The left eigenvectors for the Alfven waves read as
{%
\footnotesize
\begin{equation}
  \label{eq:39}
  \begin{split}
    \VEC l_3 & = \frac{\sign (B_1)}{2}\,\bigl( -\sign (B_1) v -
    B_2\,,\,\, 0\,,\, \sign  (B_1)\,,\,
    -\,\frac{(B_1-\sign (B_1)u)(B_2 + \sign (B_1)v)}{(u-\sign (B_1)
      B_1)^2 - c_\psi^2}\,,\, 1\,,\, -\,\frac{v+\sign (B_1)B_2}{(u-\sign (B_1)
      B_1)^2 - c_\psi^2}\bigr)^T\;, \\
    \VEC l_4 & = \frac{-\sign (B_1)}{2}\,\bigl( \sign (B_1) v - B_2\,,\,\,
    0\,,\, -\sign  (B_1)\,,\,
    -\,\frac{(B_1+\sign (B_1)u)(B_2 - \sign (B_1)v)}{(u+\sign (B_1)
      B_1)^2 - c_\psi^2}\,,\, 1\,,\, -\,\frac{v-\sign (B_1)B_2}{(u+\sign (B_1)
      B_1)^2 - c_\psi^2}\bigr)^T\;.
  \end{split}
\end{equation}
}%
The left eigenvectors for the magneto-gravitational waves are
\begin{equation}
  \label{eq:40}
  \begin{split}
    \VEC l_2 & = \frac{1}{2c_g}\,\bigl(
    u+c_g\,,\,-1\,,\,0\,,\,-\,\frac{2B_1\,(u-c_g)}{c_\psi^2 -
      (u-c_g)^2}\,,\,0\,,\,-\,\frac{2B_1}{c_\psi^2 - (u-c_g)^2} \bigr)\;, \\
    \VEC l_5 & = \frac{1}{2c_g}\,\bigl(
    -(u-c_g)\,,\,1\,,\,0\,,\,\frac{2B_1\,(u+c_g)}{c_\psi^2 -
      (u+c_g)^2}\,,\,0\,,\,\frac{2B_1}{c_\psi^2 - (u+c_g)^2} \bigr)\;.
  \end{split}
\end{equation}
The main differences to the corresponding eigenvectors for the
standard system (cf.\ equation~\eqref{eq:153}) are the changes in the
fourth component and the new sixth component.

\subsection{Non-conservative equations}
\label{sec:nonc-equat}

\subsubsection{The Powell correction}
\label{sec:powell-correction}

The above systems lack Galilean invariance. This can be restored by
giving up conservation, e.\,g.\ for the standard
system~\eqref{eq-smhd:65}--~\eqref{eq:2} by replacing the right hand
side of~(\ref{eq-smhd:65})--(\ref{eq:2}) with
\begin{equation}
  \label{eq:4}
  - (0,\: (\nabla \cdot (h\VEC B))\,\VEC B,\: (\nabla \cdot (h\VEC B))\,\VEC v)^T 
\end{equation}
which results in the so called Powell-correction.

Thus, the system matrix in the first space direction reads as
\begin{equation}
  \label{eq:5}
    \MAT A_\text{Powell} =
  \begin{pmatrix}
    0 & 1 & 0 & 0 & 0 \\
    -u^2 + B_1^2 + gh & 2u & 0 & -B_1 & 0 \\
    -uv + B_1 B_2 & v & u & 0 & -B_1 \\
    0 & 0 & 0 & u & 0 \\
    vB_1 - uB_2 & B_2 & -B_1 & 0 & u
  \end{pmatrix}\;. 
\end{equation}
Now the third Eigenvalue is no longer zero but
\begin{equation}
  \label{eq:6}
  \lambda_3 = u\;.
\end{equation}
The right eigenvectors~\(\VEC r_1\),~\(\VEC r_2\),~\(\VEC r_4\),
and~\(\VEC r_5\) remain unchanged, whereas the right eigenvector of
the modified eigenvalue reads as
\begin{equation}
  \label{eq:7}
  \VEC r_3 =
  \begin{pmatrix}
    B_1 \\ u B_1 \\ v B_1 \\ c_g^2 \\ B_2 B_1
  \end{pmatrix}\;, 
\end{equation}
which is simpler than for the conservative system. Furthermore it
guarantees hyperbolicity of the system. For the left eigenvectors, the
only change is in the fourth component, the component connected
to~\(B_1\):
\begin{equation}
  \label{eq:8}
  \begin{split}
    \VEC l_1 & = \frac{1}{2c_{g}}\, \Bigl(u+c_{g}\,,\, -1\,,\, 0\,,\,
    -\frac{B_1}{c_{g}}\,,\, 0 \Bigr)      \\
    \VEC l_2 & = \frac{1}{2}\,\Bigl( -\sign(B_1)B_2 - v\,,\, 0\,,\, 1\,,\,
    0\,,\, \sign (B_1)  \Bigr)  \\
    \VEC l_3 & = \Bigl( 0\,,\, 0\,,\, 0\,,\,
    \frac{1}{c_g^2}\,,\, 0 \Bigr)   \\
    \VEC l_4 & = -\,\frac{1}{2}\,\Bigl( \sign(B_1)B_2 - v\,,\, 0\,,\, 1\,,\,
    0\,,\, -\sign (B_1)   \Bigr)  \\
    \VEC l_5 & = -\,\frac{1}{2c_{g}}\, \Bigl(u-c_{g}\,,\, -1\,,\, 0\,,\,
    \frac{B_1}{c_{g}}\,,\, 0 \Bigr)   
  \end{split}
\end{equation}

\subsubsection{Combining Powell- and GLM-correction}
\label{sec:equations-with-glm}

In order to apply the GLM divergence correction to the Powell
system~\eqref{eq-smhd:65}--~\eqref{eq:2}, we have to add, in the
momentum equation~(\ref{eq:1}), the gradient of~\(h\psi\) to the flux,
and the the conservation equation for~\(h\psi\)
\begin{equation}
  \label{eq:9}
  (h\psi)_t + \nabla \cdot (h\psi \VEC v + c^2h\VEC B) = 0\;,
\end{equation}
where \(c\) is the artificial (constant) wave speed at which the
magnetic charges are radiated. Equation~(\ref{eq:9}) means
that~\(\psi\) is transported with the flow, and that it is
conserved. The flux Jacobian for the first space direction reads as
\begin{equation}
  \label{eq:10}
  \MAT A_\text{PGLM-naive} =
  \begin{pmatrix}
        0 & 1 & 0 & 0 & 0 & 0 \\
    -u^2 + B_1^2 + gh & 2u & 0 & -2B_1 & 0 & 0 \\
    -uv + B_1 B_2 & v & u & -B_2 & -B_1 & 0 \\
    0 & 0 & 0 & 0 & 0 & 1 \\
    vB_1 - uB_2 & B_2 & -B_1 & -v & u & 0\\
    u \psi & \psi & 0 & c^2 & 0 & u
  \end{pmatrix}\;. 
\end{equation}
The eigensystem of this matrix is rather awkward. This is mainly due
to the transport of the magnetic potential with the flow in
equation~\eqref{eq:9}. If we would omit the transport term~\(h\psi
\VEC v\) in the flux term in equation~\eqref{eq:9}, at least the
eigenvalues would be as expected, and the eigenvectors less
complicated, but this is not discussed further here.


If we make the system Galilean invariant, that means, if we subtract
\begin{equation}
  \label{eq:11}
  (0,\: (\nabla \cdot (h\VEC B))\,\VEC B,\: (\nabla \cdot (h\VEC
  B))\,\VEC v,\: -h\psi \nabla \cdot \VEC v)^T 
\end{equation}
from the right hand side of the conservation system, the system matrix
for the first space direction becomes
\begin{equation}
  \label{eq:44}
  \MAT A_\text{PGLM} =
  \begin{pmatrix}
        0 & 1 & 0 & 0 & 0 & 0 \\
    -u^2 + B_1^2 + gh & 2u & 0 & -B_1 & 0 & 0 \\
    -uv + B_1 B_2 & v & u & 0 & -B_1 & 0 \\
    0 & 0 & 0 & u & 0 & 1 \\
    vB_1 - uB_2 & B_2 & -B_1 & 0 & u & 0\\
    0 & 0 & 0 & c^2 & 0 & u
  \end{pmatrix}\;. 
\end{equation}
The eigenvalues
now are
\begin{equation}
  \label{eq:12}
  \lambda_1 = u-c\;,\quad \lambda_2\;, = u-c_g\;, \quad \lambda_3 =
  u-\abs{B_1}\;,\quad \lambda_4 = u+\abs{B_1}\;,\quad \lambda_5\;, =
  u+c_g\;, \quad  \lambda_6 = u+c\;.
\end{equation}
The right eigenvectors are much simpler than for GLM without Powell:
\begin{align}
  \label{eq:13}
  \VEC r_2 & =
  \begin{pmatrix}
    1 \\ u-c_{g} \\ v \\ 0 \\ B_2 \\ 0
  \end{pmatrix}\;,& 
  \VEC r_5 & =
  \begin{pmatrix}
    1 \\ u+c_{g} \\ v \\ 0 \\ B_2 \\ 0
  \end{pmatrix} \\ 
  \intertext{for the magneto-gravitational waves, and}
  \VEC r_3 & =
  \begin{pmatrix}
    0 \\ 0 \\ 1 \\ 0 \\ \sign (B_1) \\ 0
  \end{pmatrix}\;,& 
  \VEC r_4 & =
  \begin{pmatrix}
    0 \\ 0 \\ 1 \\ 0 \\ - \sign (B_1) \\ 0\;,
  \end{pmatrix} \\ 
  \intertext{for the Alfven waves,which means that compared to the
    conservative system without GLM-correction only a sixth component
    with value~\(0\) is appended. For the waves advecting and
    radiating away the divergence errors, the right eigenvectors are}
  \VEC r_1 & =
  \begin{pmatrix}
    -B_1 \\ -B_1\,(u-c) \\ -B_1 v \\ c^2 - c_g^2 \\ -B_1 B_2 \\
    -c\,(c^2-c_g^2) 
  \end{pmatrix}\;, &
  \VEC r_6 & =
  \begin{pmatrix}
    B_1 \\ B_1\,(u+c) \\ B_1 v \\ -(c^2 - c_g^2) \\ B_1 B_2 \\
    -c\,(c^2-c_g^2) 
  \end{pmatrix}\;. &
\end{align}
The first, third and fifth component coincide with the according
eigenvector for the Powell system without GLM. 
The left eigenvalues are
\begin{equation}
  \label{eq:14}
  \begin{split}
    \VEC l_1 & = \frac{1}{2c\,(c^2-c_g^2)} \Bigl( 0\,,\, 0\,,\, 0\,,\,
    c\,,\, 0\,,\, -1\Bigr)   \\
    \VEC l_2 & = \frac{1}{2c_{g}}\, \Bigl(u+c_{g}\,,\, -1\,,\, 0\,,\,
    \frac{c_gB_1}{c^2 - c_{g}^2}\,,\, 0\,,\, -\frac{B_1}{c^2-c_g^2}
    \Bigr)      \\ 
    \VEC l_3 & = \frac{1}{2}\,\Bigl( -\sign(B_1)B_2 - v\,,\, 0\,,\, 1\,,\,
    0\,,\, \sign (B_1)\,,\, 0  \Bigr)  \\
    \VEC l_4 & = -\,\frac{1}{2}\,\Bigl( \sign(B_1)B_2 - v\,,\, 0\,,\, 1\,,\,
    0\,,\, -\sign (B_1)\,,\, 0   \Bigr)  \\
    \VEC l_5 & = -\,\frac{1}{2c_{g}}\, \Bigl(u-c_{g}\,,\, -1\,,\, 0\,,\,
    -\frac{c_gB_1}{c^2 - c_{g}^2}\,,\, 0\,,\, -\frac{B_1}{c^2-c_g^2}
    \Bigr)   \\   
    \VEC l_1 & = \frac{1}{2c\,(c^2-c_g^2)} \Bigl( 0\,,\, 0\,,\, 0\,,\,
    c\,,\, 0\,,\, -1\Bigr)\;.   
  \end{split}
\end{equation}

In order to avoid resonance, we have to make sure that~\(c\neq
c_g\). Otherwise, the first and second as well as the fifth and sixth
eigenvectors would coincide. For practical use, it is desirable to
have~\(c > c_g\) so that~\(\lambda_1 < \lambda_2\) and~\(\lambda_6 >
\lambda_5\), which ensures that the waves transporting the unphysical
magnetic charge interfere with the other waves as little as possible.

Since the first and the sixth wave do not transport any physical feature,
but are intended to transport divergence errors out of the
computational domain, there is no real need to make them Galilean
invariant. In fact, we could keep the eigenvectors and change the
eigenvalues to
\begin{equation}
  \label{eq:30}
  \lambda_1 = -c\;,\quad \lambda_2\;, = u-c_g\;, \quad \lambda_3 =
  u-\abs{B_1}\;,\quad \lambda_4 = u+\abs{B_1}\;,\quad \lambda_5\;, =
  u+c_g\;, \quad  \lambda_6 = c\;.
\end{equation}
If the artificial wave speed~\(c\) is larger than any physical wave
speed, we not only avoid resonance, but also make sure that the errors
are quickly transported out of the computational domain. The
corresponding system matrix for the first space direction then reads
as 
\begin{equation}
  \label{eq:15}
  \MAT A_\text{PGLM-nG} =      
  \begin{pmatrix}
        0 & 1 & 0 & \frac{u B_1}{c^2-c_g^2} & 0 & 0 \\
    -u^2 + B_1^2 + gh & 2u & 0 & -B_1 + \frac{u^2 B_1}{c^2-c_g^2} & 0 & \frac{u B_1}{c^2-c_g^2} \\
    -uv + B_1 B_2 & v & u & \frac{u v B_1}{c^2-c_g^2} & -B_1 & 0 \\
    0 & 0 & 0 & 0 & 0 & 1 \\
    vB_1 - uB_2 & B_2 & -B_1 & \frac{u B_1 B_2}{c^2-c_g^2} & u & 0\\
    0 & 0 & 0 & c^2 & 0 & 0
  \end{pmatrix}\;, 
\end{equation}
where the additional \emph{nG} in the name of the matrix refers to the
fact that the system is \textbf{n}ot \textbf{G}alilean invariant.

In the following sections, we introduce Roe-type discretizations for
these systems and test them using two prototypical cases.

\section{Roe matrices for SMHD}
\label{sec:roe-matrices-smhd}

In order to arrive at numerical schemes with a high resolution of
physical phenomena and a low impact of divergence errors on the
solution, it is advantageous to apply Roe-type schemes since they
allow us to adjust the numerical viscosity for all waves
separately. The first steps involve finding Roe-linearizations for
above systems, which is not a trivial task since most of the systems
are not even conservative.

\subsection{Roe mean values and Roe matrix for the standard equations}
\label{sec:roe-mean-values}

In~\cite{roe-orig}, Roe introduces the concept of consistent local
linearizations. If we consider a Riemann problem with left and right
states~\(\VEC q_l\) and~\(\VEC q_r\) and let \(\VEC f(\VEC q)\) denote
the conservative flux and~\(\MAT A(\VEC q)\) the corresponding flux
Jacobian, then~\(\tilde{\MAT A} (\VEC q_l, \VEC q_r)\) is called a Roe
matrix, i.\,e. a matrix which defines a consistent local
linearization for the given Riemann data, if it satisfies
\begin{gather}
  \label{eq:45}
  \VEC f (\VEC q_r) - \VEC f (\VEC q_l) = \tilde{\MAT A} (\VEC q_l, \VEC
  q_r)\,(\VEC q_r - \VEC q_l)\;,  \\
  \label{eq:47}
  \tilde{\MAT A} (\VEC q_l, \VEC q_r) \to \MAT A(\VEC q) \qquad
  \text{for}\quad (\VEC q_l, \VEC q_r) \to (\VEC q, \VEC q)\;,
   \\ 
  \label{eq:51}
  \tilde{\MAT A} (\VEC q_l, \VEC q_r)\ \text{is diagonalizable for
    all}\ \VEC q_l, \VEC q_r\;.  
\end{gather}
If there exists a single
state~\(\tilde{\VEC q}=\tilde{\VEC q}(\VEC q_l, \VEC q_r)\) with
\begin{equation}
  \label{eq:46}
  \tilde{\MAT A} (\VEC q_l, \VEC q_r) = \MAT A(\tilde{\VEC q})\;, 
\end{equation}
then this state is called a \emph{Roe mean value} for \(\VEC q_l, \VEC
q_r\). For resonant hyperbolic systems, condition~\eqref{eq:51} has to
be relaxed since~\(\MAT A_\text{standard}(\VEC q)\) itself is not
always diagonalizable, but still has real eigenvalues.

For the conservative SMHD system~\eqref{eq-smhd:65}--~\eqref{eq:2}, it
is known~\cite{rossmanith} that~\(\tilde{\VEC q}\) defined by
\begin{equation}
  \label{eq:110}
  \begin{split}
    \tilde h & = \frac{1}{2}\,(h_l + h_r)\;, \\
    \tilde u & = \frac{\sqrt{h_l} u_l + \sqrt{h_r}
      u_r}{\sqrt{h_l} + \sqrt{h_r}}\;, \\ 
    \tilde v & = \frac{\sqrt{h_l} v_l + \sqrt{h_r}
      v_r}{\sqrt{h_l} + \sqrt{h_r}}\;, \\ 
    \tilde {B}_1 & = \frac{\sqrt{h_l} {B_1}_l + \sqrt{h_r}
      {B_1}_r}{\sqrt{h_l} + \sqrt{h_r}}\;, \\
    \tilde {B}_2 & = \frac{\sqrt{h_l} {B_2}_l + \sqrt{h_r}
      {B_2}_r}{\sqrt{h_l} + \sqrt{h_r}}
  \end{split}
\end{equation}
is a Roe mean value, i.\,e.~\(\MAT A (\tilde{\VEC q})\) is a
Roe-matrix as discussed above: although not always
diagonalizable, it still has real eigenvalues. 

If we augment~\(\tilde{\VEC q}\) by an arbitrary mean~\(\tilde \psi\)
of~\(\psi_l\) and~\(\psi_r\), then \(\MAT A_\text{GLM}(\tilde{\VEC
  q})\) becomes a Roe matrix for the GLM-system. This becomes clear
when considering that~\(\MAT A_\text{GLM}\) (cf.\
equation~\eqref{eq:35}) does not depend on~\(\psi\), and the
additional non-zero entries compared to~\(\MAT A_\text{standard}\)
(cf.\ equation~\eqref{eq:3}) are mere constants. If~\(c_\psi\) is
chosen large enough, it is even guaranteed that~\(\MAT
A_\text{GLM}(\tilde{\VEC q})\) satisfies all conditions for a Roe
matrix, including diagonalizability.

\subsection{Roe matrices for Powell with and without GLM}
\label{sec:roe-matrices-powell}

In order to be able to generalize the Roe scheme to non-conservative
systems, one has to replace conditions~\eqref{eq:45}--~\eqref{eq:51}
by conditions which are independent from the flux function. As
Toumi~\cite{Toumi1992360} pointed out, it is sufficient to
replace~\eqref{eq:45}. It is well known that (for sufficiently
smooth~\(\VEC f\))
\begin{equation}
  \label{eq:17}
  \VEC f (\VEC q_r) - \VEC f (\VEC q_l) = \int_0^1 A(\VEC \phi(s))\VEC
  \phi_s(s)\: \d{s}
\end{equation}
for any path~\(\VEC \phi: [0,1] \to \mathbb R^d\) with~\(\VEC \phi(0) =
\VEC q_l\), \(\VEC \phi(1) = \VEC q_r\), and~\(d\) being the dimension
of the state space. Thus, a suitable replacement for~\eqref{eq:45} has
to take on the form 
\begin{equation}
  \label{eq:16}
  \int_0^1 A(\VEC \phi(s))\VEC
  \phi_s(s)\: \d{s}
  = \tilde{\MAT A} (\VEC q_l, \VEC
  q_r)\,(\VEC q_r - \VEC q_l)\;.
\end{equation}
The main difference between the conservative and the non-conservative
case is that in the latter different choices of~\(\phi\) might lead to
different Roe matrices. 

Considering Toumi~\cite{Toumi1992360}, such a Roe matrix can be
obtained by generalizing the approach in~\cite{roe-orig}. For given
left and right states~\(\VEC q_l\) and~\(\VEC q_r\),
parametrize~\(\VEC q\) as~\(\VEC q = \VEC q(\VEC w_l + (1-s)\VEC
w_r)\) with~\(\VEC q (\VEC w_l) = \VEC q_l\) and~\(\VEC q (\VEC w_r) =
\VEC q_r\). Thus,~\(\VEC \phi\) becomes~\(\VEC \phi(s) = \VEC q(\VEC
w_l + (1-s)\VEC w_r)\).  With that, the integral on the left-hand side
of equation~\eqref{eq:16} becomes
\begin{equation}
  \label{eq:19}
  \int_0^1 A(\VEC \phi(s))\VEC \phi_s(s)\: \d{s}
  = \int_0^1 \MAT A\bigl(\VEC q(\VEC w_l + (1-s)\VEC w_r)\bigr)\cdot
  \MAT A_0(\VEC w_l + (1-s)\VEC w_r)\: \d{s} \cdot (\VEC w_r - \VEC w_l)\;,
\end{equation}
where~\(\MAT A_0\) is the Jacobian of~\(\VEC q\) with respect
to~\(\VEC w\).  On the other hand, for~\(\MAT A_0\) the mean value
theorem implies
\begin{equation}
  \label{eq:20}
  \VEC q_r - \VEC q_l = \int_0^1 \MAT A_0(s\,\VEC w_r + (1-s)\,\VEC w_l)\:
  \d{s} \cdot (\VEC w_r - \VEC w_l)\;.
\end{equation}
The integral on the right-hand side of equation~\eqref{eq:20} is non-singular for any reasonable parametrization of~\(\VEC q\), thus
\begin{equation*}
  \tilde{\MAT A} (\VEC q_l, \VEC q_r) = 
  \int_0^1 \MAT A\bigl(\VEC q(s\,\VEC w_l + (1-s)\,\VEC w_r)\bigr)\cdot
  \MAT A_0(\VEC w_r + (1-s)\VEC w_l)\: \d{s}\,\cdot   
  \Bigl(\int_0^1 \MAT A_0(\VEC w_l + (1-s)\VEC w_r)\: \d{s}  \Bigr)^{-1}
\end{equation*}
is a Roe matrix with respect to the path~\(\VEC \phi(s) = \VEC q(s\,\VEC
w_r + (1-s)\,\VEC w_l)\).

For the standard SMHD system~\eqref{eq-smhd:65}--~\eqref{eq:2} the usual
choice is
\begin{equation}
  \label{eq:21}
  \VEC w := (\sqrt{h},\sqrt{h}u,\sqrt{h}v,\sqrt{h}B_1,\sqrt{h}B_2)^T
\end{equation}
and, thus,
\begin{equation}
  \label{eq:22}
  \VEC q(\VEC w) = \sqrt{h}\,\VEC w\;.
\end{equation}
This leads to the Roe matrix given in~\cite{rossmanith}, which is just
the matrix~ \(\MAT A (\tilde{\VEC q})\) described in the previous
section.  

If the fourth column in the system matrix~\(\MAT A_\text{Powell}\) for
the system with the Powell correction (equation~\eqref{eq:5}) had only
the diagonal entry, the Roe matrix given our choice
for~\(\VEC w\) would simply be the matrix~\(\MAT
A_\text{Powell}(\tilde{\VEC q})\). But the entry in the second row
causes significant
changes  in the first column of the Roe
matrix: {%
  \setlength{\extrarowheight}{.8em}
\begin{equation}
  \label{eq:23}
      A_1 =
  \begin{pmatrix}
    0 & 1 & 0 & 0 & 0 \\
    -\tilde u^2 + \frac{1}{2}\,\Bigl[\tilde B_1^2 + \widehat{B_1^2}\Bigr] + g\tilde h & 2\tilde u & 0 & -\tilde B_1 & 0 \\
    -\tilde u\tilde v + \frac{1}{2}\,\Bigl[\tilde B_1 \tilde B_2 +
    \widehat{B_1 B_2}\Bigr] & \tilde v & \tilde u & 0 & -\tilde B_1 \\
    \frac{1}{2}\,\Bigl[\widehat{u B_1} - \tilde u \tilde B_1\Bigr] & 0 & 0 & \tilde u & 0 \\
    \frac{1}{2}\,\Bigl[\tilde v \tilde B_1 + \widehat{v B_1}\Bigr] - \tilde u \tilde B_2 & \tilde B_2 & -\tilde B_1 & 0 & \tilde u
  \end{pmatrix}\;, 
\end{equation}
}%
where, e.\,g., the hatted term in the last row is defined as
\begin{equation}
  \label{eq:24}
  \widehat{v B_1} = \frac{\overline{\bigl(\frac{w_3
        w_4}{w_1}\bigr)}}{\overline w_1}
\end{equation}
with
\begin{equation}
  \label{eq:25}
  \overline{\bigl(\frac{w_3 w_4}{w_1}\bigr)}:=\int_0^1 \frac{\bigl(s
    \,{w_3}_r + (1-s)\,{w_3}_l\bigr)\bigl(s
    \,{w_4}_r + (1-s)\,{w_4}_l\bigr)}{\bigl(s
    \,{w_1}_r + (1-s)\,{w_1}_l\bigr)}\:\d{s}
\end{equation}
and
\begin{equation}
  \label{eq:26}
  \overline w_1 :=\int_0^1 s\,{w_1}_r + (1-s)\,{w_1}_l\:\d{s} =
  \frac{{w_1}_r +{w_1}_l}{2}\;, 
\end{equation}
where the~\(w_k\) are the components of the parameter vector~\(\VEC
w\) as defined in equation~\eqref{eq:21}. (The other hatted terms are
defined likewise.)The integral in equation~\eqref{eq:25} can be solved
explicitly, but the solution involves many terms, some of them
logarithmic. This makes the computation rather time
consuming. Furthermore, the fill-in in the fourth row of the first
column makes the computation of the eigensystem rather awkward and
even more time consuming. Thus, we decide to drop this approach.

In order to avoid unphysical solutions, we have to recall the
derivation of the system. As long as there is no magnetic charge or
magnetic current, both the conservative standard system and the
Galilean invariant Powell system yield the same analytical
solutions. Thus, it seems sufficient to extend the Roe matrix for the
conservative system in the same way as we did for the analytic
case. For our numerical computations, we employ the local
linearizations~\(\MAT A_\text{Powell}(\tilde{\VEC
  q})\)~(equation~\eqref{eq:5}) and~\(\MAT A_\text{PGLM}(\tilde{\VEC
  q})\)~(equation~\eqref{eq:44}) for the Powell case and the
Powell-GLM case respectively. This ensures that in the absence of
magnetic charges---which are just numerical artifacts---we obtain
physical solutions. By the construction of the Powell and the
Powell-GLM systems, we expect these numerical artifacts to be
transported and/or radiated away from the computational domain.

In more detail: If we write the system matrix~\(\MAT A_\text{Powell}({\VEC q})\)
of the Powell system as
\begin{equation}
  \label{eq:18}
  \MAT A_\text{Powell}({\VEC q}) =  \MAT A_\text{standard}({\VEC q}) +  \MAT A_\text{corr}({\VEC
    q}) 
\end{equation}
and perform the above process for both parts separately, we obtain
\begin{equation}
  \label{eq:27}
  \tilde{\MAT A}_\text{Powell} (\VEC q_l, \VEC q_r) = \MAT
  A_\text{standard}(\tilde{\VEC q}) +  \tilde{\MAT A}_\text{corr}
  (\VEC q_l, \VEC q_r)
\end{equation}
with
{%
\setlength{\extrarowheight}{.6em}
\begin{equation}
  \label{eq:28}
  \tilde{\MAT A}_\text{corr} (\VEC q_l, \VEC q_r) 
  = \MAT A_\text{corr}(\tilde{\VEC q})
  + 
  \begin{pmatrix}
    0 & 0 & 0 & 0 & 0 \\
    \frac{1}{2} \Bigl(\widehat{B_1^2} - \tilde B_1^2 \Bigr) & 0 & 0 &
    0 & 0 \\
    \frac{1}{2} \Bigl(\widehat{B_1 B_2} - \tilde B_1 \tilde B_2 \Bigr)
    & 0 & 0 & 0 & 0 \\
    \frac{1}{2} \Bigl(\widehat{u B_1} - \tilde u \tilde B_1  \Bigr) &
    0 & 0 & 0 & 0 \\
    \frac{1}{2} \Bigl(\widehat{B_1 B_2} - \tilde v \tilde B_1  \Bigr)
    & 0 & 0 & 0 & 0 
  \end{pmatrix}\,.
\end{equation}
}%
The only remaining terms in the second matrix on the right-hand side
result from the different averaging strategies.


\section{Numerical tests}
\label{sec:numerical-tests}

All tests are done with clawpack~\cite{clawpack}, an implementation of
the wave propagation method, which is based on fluctuation
splitting. Thus, it can handle non-conservative systems in a natural
way. Since in practice, most computations are done with second or even
higher order, we present results only for second order. As limiters,
we employ Superpower~\cite{limiter} on the magnetogravitational
waves and CFL-superbee on all other waves. The resonant wave in the
standard model is computed with first order. The implementation
follows~\cite{limiter} and~\cite{kemm-santiago}, which explain how to
apply the Jeng-Payne correction for spatial varying CFL-numbers to
arbitrary TVD-limiters.

To avoid expansion shocks, we employ the Harten entropy
fix\footnote{Not to be confused with the Harten-Hyman entropy
  fix~\cite{harten-hyman}}~\cite{harten-tvd}. For this, a
parameter~\(\delta > 0\) has to be chosen. For values smaller
than~\(\delta\), the numerical viscosity on the wave is modified and,
finally, bounded from below by~\(\delta/2\) according to
\begin{equation}
  \label{eq:42}
    \phi(\lambda) =
  \begin{cases}
    \abs{\tilde\lambda} & \text{if}~\abs{\tilde\lambda} \geq \delta\;,\\
    (\tilde{\lambda}^2+\delta^2)/(2\delta) &
    \text{if}~\abs{\tilde\lambda} < \delta\;, 
  \end{cases}
\end{equation}
where~\(\phi(\tilde\lambda)\) denotes the viscosity on the wave
corresponding to the eigenvalue~\(\tilde\lambda\) of the Roe
matrix. If not stated otherwise, we use~\(\delta=10^{-8}\) on all
waves.  For the standard model, we employ as the Harten parameter for
the resonant wave
\begin{equation}
  \label{eq:41}
  \delta_\text{resonant} = 2\,\max{\abs{\tilde u - \tilde
    c_g},\abs{\tilde u + \tilde c_g}} =: 2\,s_\text{max}
\end{equation}
as suggested in~\cite{orig-div,habil-kemm}. Note that for the standard
GLM (cf.\ equation~\eqref{eq:35}) and the system determined by
equation~\eqref{eq:15}, the application of the Harten entropy fix on
the artificial waves is pointless. Their speed is
always~\(c_\psi\). Furthermore,~\(c_\psi\) is chosen to be the highest
wave speed and, thus, determines the time step. Therefore, raising the
viscosity on these waves would give no advantage over choosing the
corresponding higher wave speed for~\(c_\psi\). Note that
wherever~\(\phi(\tilde\lambda)\neq\abs{\tilde\lambda}\), the accuracy
of the scheme drops to first order. This is due to the algebraic
limiting used in clawpack. As we already pointed out
in~\cite{habil-kemm}, algebraic limiting is formulated as a correction
to a fixed first order scheme, in clawpack the wave-wise standard
upwind scheme. If the underlying first order scheme is changed, the
correction, in general, cannot lead to second order of the resulting
scheme.

The artificial speed~\(c_\psi\) is stated as multiples
of~\(s_\text{max}\), which, as indicated in equation~\eqref{eq:41},
represents the highest physical wave speed in the problem. This is
done for each time step separately, i.\,e., before the time step is
performed,~\(c_\psi\) is set according to the wave speeds occurring in
the last time step.

In order to assess above methods, we consider the
normalized~\(\mathscr L^\infty\)-,~\(\mathscr L^2\)-,~\(\mathscr
L^1\)-norms. We also show some pictures of the height and the
distribution of the divergence errors to illustrate the effects
detected in the plots of the norms.
We employ two prototypical test cases: (1) a generalized radial dam
break as an example, where plasma will remain inside the boundary of
the computational domain for a long time. This is a challenge
especially for the pure Powell correction. (2) the de~Sterck test as
an example with a high flow speed in \(x\)-direction. This is a
challenge especially for the Powell-GLM system since it might happen
that~\(u-c\) is positive or even vanishes, i.\,e., the divergence
errors could stay in the computational domain for a prolonged time.

\subsection{Generalized radial dam break}
\label{sec:separ-cond-fluids}

We start with the generalized dam break problem, which was introduced
as a test case by Quamar et~el.~\cite{Qamar2006132}. The initial state
is a situation at rest. In addition,~\(h\VEC B\) is constant all over
the computational domain and, thus, divergence-free.  In the center of
the computational domain is a cylindrical region of radius \(0.1\)
with increased height. The initial data for the fluids in the circular
and outside regions are:
\begin{align}
  \label{eq:29}
  h_\text{in} & = 10\;, & u_\text{in} & = 0\;, & v_\text{in} & = 0\;,
  & {B_1}_\text{in} & = 0.1\;, & {B_2}_\text{in} & = 0\;, \\
  h_\text{out} & = 1\;, & u_\text{out} & = 0\;, & v_\text{out} & = 0\;,
  & {B_1}_\text{out} & = 1\;, & {B_2}_\text{out} & = 0\;,
\end{align}
and the gravitational constant is normalized to \(g = 1\). For the
consideration of the error norms, we compute up to~\(t=0.3\).  The
computational domain is a square region~\([-1, 1]\times [-1, 1]\) with
extrapolation boundaries. The domain is discretized with~\(300\times
300\) grid cells. 

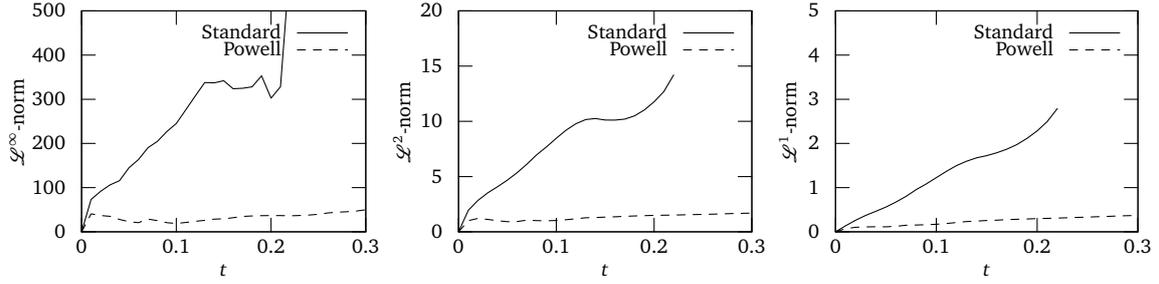
\begin{figure}
  \centering{
  \begin{tikzpicture}[gnuplot]
\tikzset{every node/.append style={scale=0.65}}
\gpmonochromelines
\gpcolor{color=gp lt color border}
\gpsetlinetype{gp lt border}
\gpsetlinewidth{1.00}
\draw[gp path] (0.980,0.640)--(1.160,0.640);
\draw[gp path] (4.719,0.640)--(4.539,0.640);
\node[gp node right] at (0.860,0.640) { 0};
\draw[gp path] (0.980,1.226)--(1.160,1.226);
\draw[gp path] (4.719,1.226)--(4.539,1.226);
\node[gp node right] at (0.860,1.226) { 100};
\draw[gp path] (0.980,1.812)--(1.160,1.812);
\draw[gp path] (4.719,1.812)--(4.539,1.812);
\node[gp node right] at (0.860,1.812) { 200};
\draw[gp path] (0.980,2.397)--(1.160,2.397);
\draw[gp path] (4.719,2.397)--(4.539,2.397);
\node[gp node right] at (0.860,2.397) { 300};
\draw[gp path] (0.980,2.983)--(1.160,2.983);
\draw[gp path] (4.719,2.983)--(4.539,2.983);
\node[gp node right] at (0.860,2.983) { 400};
\draw[gp path] (0.980,3.569)--(1.160,3.569);
\draw[gp path] (4.719,3.569)--(4.539,3.569);
\node[gp node right] at (0.860,3.569) { 500};
\draw[gp path] (0.980,0.640)--(0.980,0.820);
\draw[gp path] (0.980,3.569)--(0.980,3.389);
\node[gp node center] at (0.980,0.440) { 0};
\draw[gp path] (2.226,0.640)--(2.226,0.820);
\draw[gp path] (2.226,3.569)--(2.226,3.389);
\node[gp node center] at (2.226,0.440) { 0.1};
\draw[gp path] (3.473,0.640)--(3.473,0.820);
\draw[gp path] (3.473,3.569)--(3.473,3.389);
\node[gp node center] at (3.473,0.440) { 0.2};
\draw[gp path] (4.719,0.640)--(4.719,0.820);
\draw[gp path] (4.719,3.569)--(4.719,3.389);
\node[gp node center] at (4.719,0.440) { 0.3};
\draw[gp path] (0.980,3.569)--(0.980,0.640)--(4.719,0.640)--(4.719,3.569)--cycle;
\node[gp node center,rotate=-270] at (0.160,2.104) {\(\mathscr L^\infty\)-norm};
\node[gp node center] at (2.849,0.140) {\(t\)};
\node[gp node right] at (3.699,3.277) {Standard};
\gpcolor{color=gp lt color 0}
\gpsetlinetype{gp lt plot 0}
\draw[gp path] (3.819,3.277)--(4.479,3.277);
\draw[gp path] (0.980,0.640)--(1.105,1.069)--(1.229,1.176)--(1.354,1.262)--(1.479,1.318)%
  --(1.603,1.489)--(1.728,1.598)--(1.852,1.754)--(1.977,1.839)--(2.102,1.970)--(2.226,2.076)%
  --(2.351,2.258)--(2.476,2.441)--(2.600,2.619)--(2.725,2.615)--(2.850,2.643)--(2.974,2.537)%
  --(3.099,2.543)--(3.223,2.562)--(3.348,2.708)--(3.473,2.412)--(3.597,2.564)--(3.673,3.569);
\gpcolor{color=gp lt color border}
\node[gp node right] at (3.699,3.052) {Powell};
\gpcolor{color=gp lt color 1}
\gpsetlinetype{gp lt plot 1}
\draw[gp path] (3.819,3.052)--(4.479,3.052);
\draw[gp path] (0.980,0.640)--(1.105,0.878)--(1.229,0.856)--(1.354,0.845)--(1.479,0.804)%
  --(1.603,0.774)--(1.728,0.761)--(1.852,0.809)--(1.977,0.788)--(2.102,0.761)--(2.226,0.750)%
  --(2.351,0.764)--(2.476,0.779)--(2.600,0.792)--(2.725,0.808)--(2.850,0.812)--(2.974,0.827)%
  --(3.099,0.844)--(3.223,0.850)--(3.348,0.854)--(3.473,0.855)--(3.597,0.855)--(3.722,0.852)%
  --(3.847,0.856)--(3.971,0.862)--(4.096,0.872)--(4.220,0.890)--(4.345,0.900)--(4.470,0.906)%
  --(4.594,0.915)--(4.719,0.932);
\gpcolor{color=gp lt color border}
\gpsetlinetype{gp lt border}
\draw[gp path] (0.980,3.569)--(0.980,0.640)--(4.719,0.640)--(4.719,3.569)--cycle;
\gpdefrectangularnode{gp plot 1}{\pgfpoint{0.980cm}{0.640cm}}{\pgfpoint{4.719cm}{3.569cm}}
\draw[gp path] (5.940,0.640)--(6.120,0.640);
\draw[gp path] (9.799,0.640)--(9.619,0.640);
\node[gp node right] at (5.820,0.640) { 0};
\draw[gp path] (5.940,1.372)--(6.120,1.372);
\draw[gp path] (9.799,1.372)--(9.619,1.372);
\node[gp node right] at (5.820,1.372) { 5};
\draw[gp path] (5.940,2.105)--(6.120,2.105);
\draw[gp path] (9.799,2.105)--(9.619,2.105);
\node[gp node right] at (5.820,2.105) { 10};
\draw[gp path] (5.940,2.837)--(6.120,2.837);
\draw[gp path] (9.799,2.837)--(9.619,2.837);
\node[gp node right] at (5.820,2.837) { 15};
\draw[gp path] (5.940,3.569)--(6.120,3.569);
\draw[gp path] (9.799,3.569)--(9.619,3.569);
\node[gp node right] at (5.820,3.569) { 20};
\draw[gp path] (5.940,0.640)--(5.940,0.820);
\draw[gp path] (5.940,3.569)--(5.940,3.389);
\node[gp node center] at (5.940,0.440) { 0};
\draw[gp path] (7.226,0.640)--(7.226,0.820);
\draw[gp path] (7.226,3.569)--(7.226,3.389);
\node[gp node center] at (7.226,0.440) { 0.1};
\draw[gp path] (8.513,0.640)--(8.513,0.820);
\draw[gp path] (8.513,3.569)--(8.513,3.389);
\node[gp node center] at (8.513,0.440) { 0.2};
\draw[gp path] (9.799,0.640)--(9.799,0.820);
\draw[gp path] (9.799,3.569)--(9.799,3.389);
\node[gp node center] at (9.799,0.440) { 0.3};
\draw[gp path] (5.940,3.569)--(5.940,0.640)--(9.799,0.640)--(9.799,3.569)--cycle;
\node[gp node center,rotate=-270] at (5.240,2.104) {\(\mathscr L^2\)-norm};
\node[gp node center] at (7.869,0.140) {\(t\)};
\node[gp node right] at (8.779,3.277) {Standard};
\gpcolor{color=gp lt color 0}
\gpsetlinetype{gp lt plot 0}
\draw[gp path] (8.899,3.277)--(9.559,3.277);
\draw[gp path] (5.940,0.640)--(6.069,0.926)--(6.197,1.058)--(6.326,1.158)--(6.455,1.239)%
  --(6.583,1.326)--(6.712,1.425)--(6.840,1.538)--(6.969,1.662)--(7.098,1.767)--(7.226,1.881)%
  --(7.355,1.988)--(7.484,2.073)--(7.612,2.128)--(7.741,2.142)--(7.870,2.122)--(7.998,2.121)%
  --(8.127,2.134)--(8.255,2.177)--(8.384,2.255)--(8.513,2.364)--(8.641,2.499)--(8.770,2.719);
\gpcolor{color=gp lt color border}
\node[gp node right] at (8.779,3.052) {Powell};
\gpcolor{color=gp lt color 1}
\gpsetlinetype{gp lt plot 1}
\draw[gp path] (8.899,3.052)--(9.559,3.052);
\draw[gp path] (5.940,0.640)--(6.069,0.788)--(6.197,0.818)--(6.326,0.804)--(6.455,0.787)%
  --(6.583,0.774)--(6.712,0.776)--(6.840,0.795)--(6.969,0.788)--(7.098,0.786)--(7.226,0.790)%
  --(7.355,0.800)--(7.484,0.813)--(7.612,0.826)--(7.741,0.829)--(7.870,0.835)--(7.998,0.839)%
  --(8.127,0.845)--(8.255,0.850)--(8.384,0.854)--(8.513,0.858)--(8.641,0.860)--(8.770,0.862)%
  --(8.899,0.865)--(9.027,0.868)--(9.156,0.871)--(9.284,0.874)--(9.413,0.878)--(9.542,0.882)%
  --(9.670,0.885)--(9.799,0.888);
\gpcolor{color=gp lt color border}
\gpsetlinetype{gp lt border}
\draw[gp path] (5.940,3.569)--(5.940,0.640)--(9.799,0.640)--(9.799,3.569)--cycle;
\gpdefrectangularnode{gp plot 2}{\pgfpoint{5.940cm}{0.640cm}}{\pgfpoint{9.799cm}{3.569cm}}
\draw[gp path] (10.900,0.640)--(11.080,0.640);
\draw[gp path] (14.879,0.640)--(14.699,0.640);
\node[gp node right] at (10.780,0.640) { 0};
\draw[gp path] (10.900,1.226)--(11.080,1.226);
\draw[gp path] (14.879,1.226)--(14.699,1.226);
\node[gp node right] at (10.780,1.226) { 1};
\draw[gp path] (10.900,1.812)--(11.080,1.812);
\draw[gp path] (14.879,1.812)--(14.699,1.812);
\node[gp node right] at (10.780,1.812) { 2};
\draw[gp path] (10.900,2.397)--(11.080,2.397);
\draw[gp path] (14.879,2.397)--(14.699,2.397);
\node[gp node right] at (10.780,2.397) { 3};
\draw[gp path] (10.900,2.983)--(11.080,2.983);
\draw[gp path] (14.879,2.983)--(14.699,2.983);
\node[gp node right] at (10.780,2.983) { 4};
\draw[gp path] (10.900,3.569)--(11.080,3.569);
\draw[gp path] (14.879,3.569)--(14.699,3.569);
\node[gp node right] at (10.780,3.569) { 5};
\draw[gp path] (10.900,0.640)--(10.900,0.820);
\draw[gp path] (10.900,3.569)--(10.900,3.389);
\node[gp node center] at (10.900,0.440) { 0};
\draw[gp path] (12.226,0.640)--(12.226,0.820);
\draw[gp path] (12.226,3.569)--(12.226,3.389);
\node[gp node center] at (12.226,0.440) { 0.1};
\draw[gp path] (13.553,0.640)--(13.553,0.820);
\draw[gp path] (13.553,3.569)--(13.553,3.389);
\node[gp node center] at (13.553,0.440) { 0.2};
\draw[gp path] (14.879,0.640)--(14.879,0.820);
\draw[gp path] (14.879,3.569)--(14.879,3.389);
\node[gp node center] at (14.879,0.440) { 0.3};
\draw[gp path] (10.900,3.569)--(10.900,0.640)--(14.879,0.640)--(14.879,3.569)--cycle;
\node[gp node center,rotate=-270] at (10.320,2.104) {\(\mathscr L^1\)-norm};
\node[gp node center] at (12.889,0.140) {\(t\)};
\node[gp node right] at (13.859,3.277) {Standard};
\gpcolor{color=gp lt color 0}
\gpsetlinetype{gp lt plot 0}
\draw[gp path] (13.979,3.277)--(14.639,3.277);
\draw[gp path] (10.900,0.640)--(11.033,0.720)--(11.165,0.794)--(11.298,0.858)--(11.431,0.912)%
  --(11.563,0.968)--(11.696,1.035)--(11.828,1.110)--(11.961,1.199)--(12.094,1.275)--(12.226,1.358)%
  --(12.359,1.442)--(12.492,1.518)--(12.624,1.576)--(12.757,1.621)--(12.890,1.650)--(13.022,1.688)%
  --(13.155,1.736)--(13.287,1.797)--(13.420,1.879)--(13.553,1.978)--(13.685,2.101)--(13.818,2.273);
\gpcolor{color=gp lt color border}
\node[gp node right] at (13.859,3.052) {Powell};
\gpcolor{color=gp lt color 1}
\gpsetlinetype{gp lt plot 1}
\draw[gp path] (13.979,3.052)--(14.639,3.052);
\draw[gp path] (10.900,0.640)--(11.033,0.682)--(11.165,0.700)--(11.298,0.705)--(11.431,0.707)%
  --(11.563,0.706)--(11.696,0.713)--(11.828,0.727)--(11.961,0.730)--(12.094,0.734)--(12.226,0.740)%
  --(12.359,0.750)--(12.492,0.763)--(12.624,0.776)--(12.757,0.782)--(12.890,0.790)--(13.022,0.794)%
  --(13.155,0.802)--(13.287,0.806)--(13.420,0.810)--(13.553,0.815)--(13.685,0.818)--(13.818,0.822)%
  --(13.951,0.826)--(14.083,0.831)--(14.216,0.835)--(14.348,0.840)--(14.481,0.844)--(14.614,0.850)%
  --(14.746,0.854)--(14.879,0.855);
\gpcolor{color=gp lt color border}
\gpsetlinetype{gp lt border}
\draw[gp path] (10.900,3.569)--(10.900,0.640)--(14.879,0.640)--(14.879,3.569)--cycle;
\gpdefrectangularnode{gp plot 3}{\pgfpoint{10.900cm}{0.640cm}}{\pgfpoint{14.879cm}{3.569cm}}
\end{tikzpicture}
}
  \caption{Generalized dam break: Different norms for divergence when
    computed without GLM.}
  \label{fig:seppow}
\end{figure}
For this generalized radial dam-break experiment, we first consider
the results obtained without GLM. Both the standard and the Powell
system have in common that all magnetic monopoles stay in the
computational domain. For the standard equations since there is no
transport of magnetic monopoles at all, for the Powell system since
the flow velocity at the boundary is zero. Thus, one would expect
the~\(\mathscr L^1\)-norm of the divergence errors to be identical for
both calculations. But, as Figure~\ref{fig:seppow} shows, for the
standard system, it is considerably higher. This confirms the major
role played by resonance in the production of divergence
errors~\cite{orig-div}. Although we imposed a high viscosity on the
resonant wave, the production of magnetic charges is much higher than
for the non-resonant Powell system.
\begin{figure}
  \centering
  \includegraphics[width=.7\linewidth]{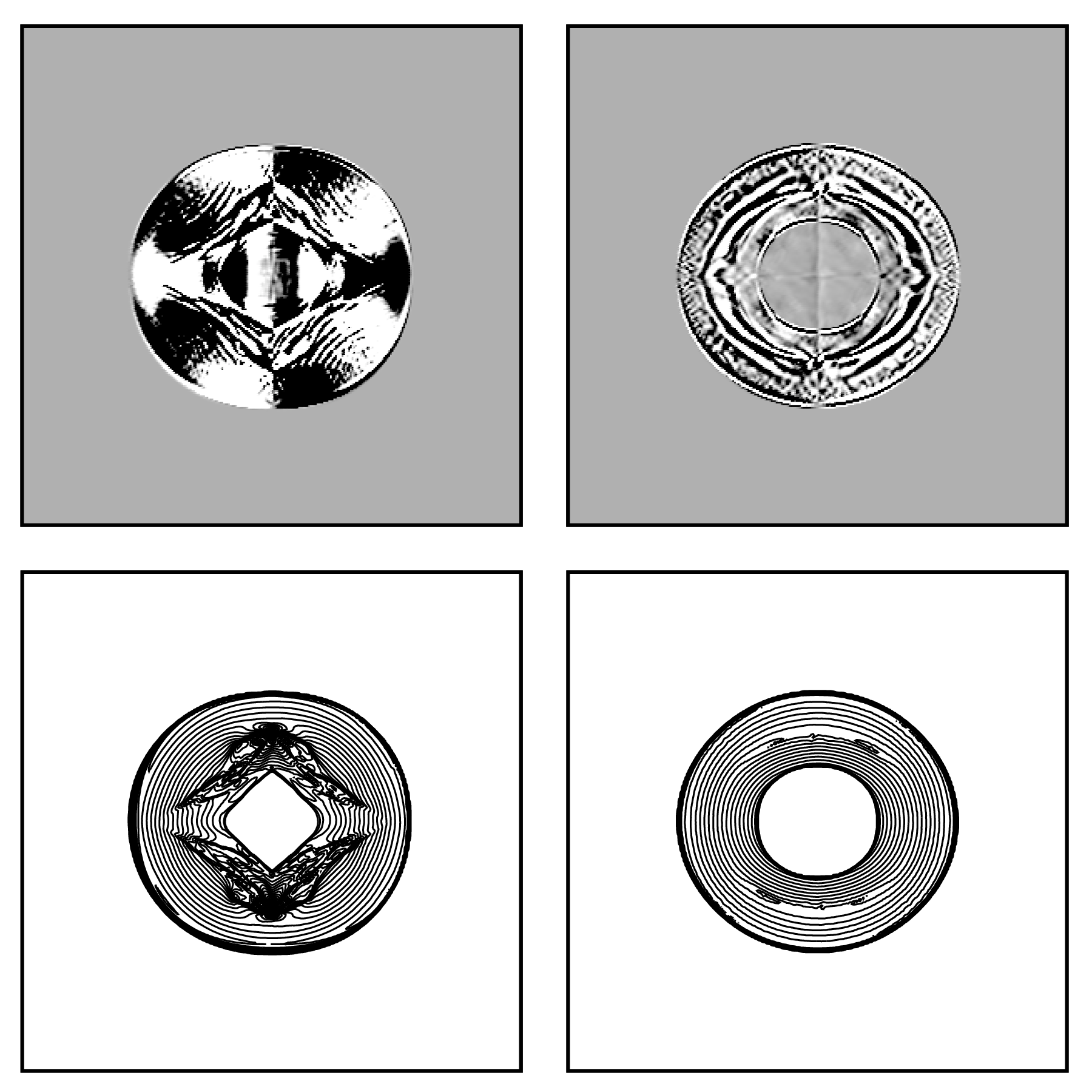}
  \caption{Generalized dam break without GLM at~\(t=0.18\): Schlieren
    type plots for divergence error (upper row) and contour plots for
    height (lower row); left without divergence correction, right with
    Powell correction.}
  \label{fig:standardfail}
\end{figure}
Figure~\ref{fig:standardfail} shows the solution of both computations
shortly before the breakdown of the standard computation due to
excessive divergence errors. For the Schlieren type plots, we
employed~\(\pm 0.3\) as a threshold for black and white.  It is
clearly seen that, while for the Powell system the height seems reasonable,
the results for the standard system are highly unphysical and thus
worthless. It can also be observed that in both cases no magnetic
monopoles are found outside of the area confined by the shock.

\begin{figure}
  \centering
  \input{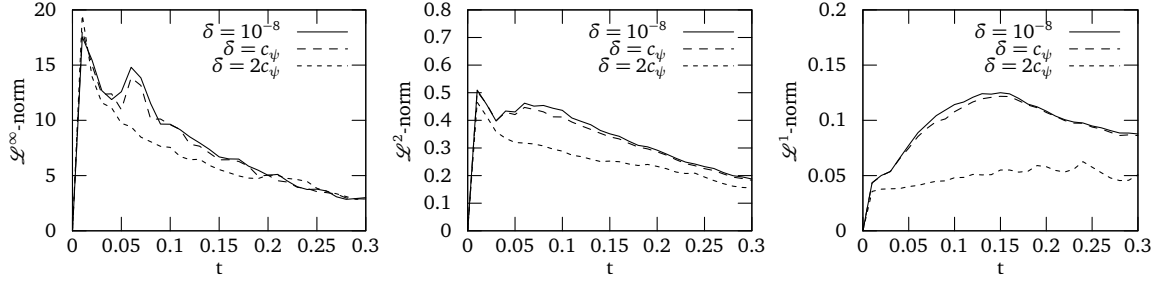}
  \caption{Generalized dam break: Powell\,+\,GLM: Comparison for different viscosities; all cases computed
    with \(c_\psi = 2s_\text{max}\).}
  \label{fig:sepvisc}
\end{figure}
\begin{figure}
  \centering
  \input{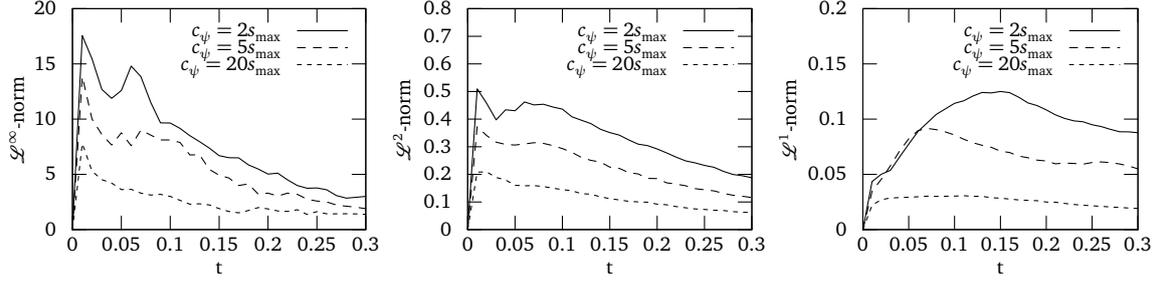}
  \caption{Generalized dam break: Powell\,+\,GLM: Comparison for different wave speeds; all cases computed
    with \(\delta = 10^{-8}\).}
  \label{fig:sepspeed}
\end{figure}
In Figures~\ref{fig:sepvisc} and~\ref{fig:sepspeed}, we show different
tests for the Powell-GLM system. All results for that system differ
from the standard and the Powell computations in one crucial point:
after some time all error norms decrease. However, this is impossible
for the Powell and the standard system. Because of the initial
conditions and the boundary conditions, divergence errors cannot be
removed from the computational domain, not even for the Powell system.

In Figure~\ref{fig:sepvisc}, we compare different choices for the
viscosity on the artificial waves. All tests are done with the
artificial speed~\(c_\psi\) set to~\(c_\psi = 2s_\text{max}\). The
viscosity is adjusted via the parameter~\(\delta\) in the Harten
entropy fix. Although the implementation of the viscosity as a split
of a single wave into two waves running at different speeds formally
would require a reduction of the time step size, we computed it using
the wave speeds of the original waves. This means that the
computational cost is identical for all tested viscosities.  As
Einfeldt~\cite{hllem} points out, the stability of the scheme does not
depend on the faster of those split waves, but on the resulting
viscosity superimposed on the original wave. As
Figure~\ref{fig:sepvisc} indicates, the computation is stable even
with~\(\delta = 2c_\psi\).  This can be inferred from
equation~\eqref{eq:42}, which implies that the difference
between~\(\abs{\tilde\lambda}\) and~\(\phi(\tilde\lambda)\) decreases
for increasing~\(\abs{\tilde\lambda}\).
Since, for
every~\(t>0\), the flow field contains supercritical regions, we can
expect (but, unfortunately, not guarantee) that the maximal value
for~\(\abs{\tilde u}\) is larger than~\(s_\text{max}/2\). Together
with the setting~\(c_\psi = 2s_\text{max}\), this means that the
maximal value for~\(\abs{\tilde u\pm c_\psi}\) (which determines the
time step size) is larger than~\(\frac{5}{4}\, c_\psi\). If we insert
this and~\(\delta = 2c_\psi\) in equation~\eqref{eq:42}, we obtain as
an upper bound for the viscosity
\begin{equation}
  \label{eq:43}
  \phi\Bigl(\frac{5}{4}\, c_\psi\Bigr) = \frac{5}{4}\, c_\psi +
  \frac{9}{64}\, c_\psi\;. 
\end{equation}
Since the time step size in our computations is computed with
CFL-number~\(0.9\) with respect to the maximal value of~\(\abs{\tilde
  u\pm c_\psi}\), this implies for the CFL-number with respect to the
viscosity an upper bound of~\(1.0125\). From our considerations, it is
clear that we usually cannot expect this upper bound to be attained at
any time~\(t>0\). In the initial state of the generalized dam break
problem, no supercritical regions are found, thus the above
assumptions are not true at~\(t=0\), and the real CFL-number might be
larger than one. This, in turn, might be the reason for the rather
high errors observed for the early time steps. Note that, since the
maximal value for~\(\abs{\tilde u\pm c_\psi}\) is bounded below
by~\(c_\psi\),~\(\delta = c_\psi\) can always be chosen without any
restrictions.

The most significant feature in Figure~\ref{fig:sepvisc} is that the
gain in quality obtained by a higher viscosity on the divergence waves
is mostly seen for the~\(\mathscr L^1\)-norm. The differences for
the~\(\mathscr L^\infty\)- and the~\(\mathscr L^2\)-norm are
small. This indicates that the latter mainly represent the production
of the divergence errors at the shock. From that, we conclude that
even without resonance, the viscosity on the divergence waves plays an
important role.

In Figure~\ref{fig:sepspeed} we show a comparison for different
choices of the artificial wave speed. The parameter~\(\delta\) is set
to its standard value~\(\delta = 10^{-8}\). Here, a difference can be
clearly seen. But one has to keep in mind that the better results are
at the price of a significantly reduced time step. Choosing~\(c_\psi =
20s_\text{max}\) instead of~\(c_\psi = 2s_\text{max}\) would reduce
the divergence errors by a factor of (roughly)~\(3\), but might reduce
the time step by a factor of up to~\(10\). As the comparison to
Figure~\ref{fig:sepvisc} shows, for~\(c_\psi =
2s_\text{max}\),~\(\delta = 2c_\psi\) yields results of a quality
almost comparable to the results for~\(c_\psi =
5s_\text{max}\),~\(\delta = 10^{-8}\). While the latter is better in
terms of the~\(\mathscr L^\infty\)-norm, the previous results in a
lower~\(\mathscr L^1\)-norm. The ~\(\mathscr L^2\)-norm is about the
same in both cases. This confirms that in general it is sufficient to
choose~\(c_\psi = 2s_\text{max}\),~\(\delta = 2c_\psi\), which allows
a higher time step size than~\(c_\psi = 5s_\text{max}\),~\(\delta =
10^{-8}\).
\begin{figure}
  \centering
  \includegraphics[width=.7\linewidth]{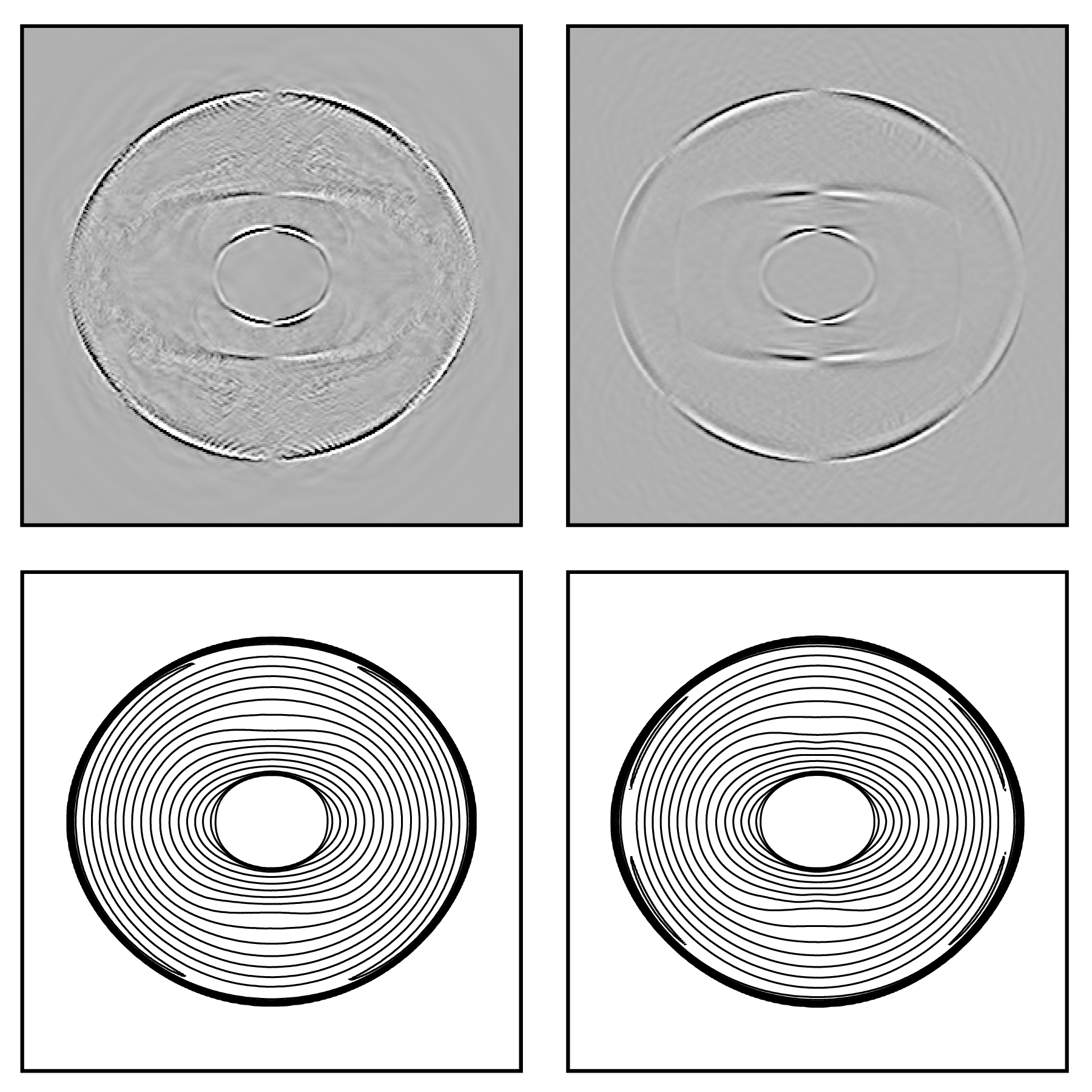}
  \caption{Generalized dam break with Powell+GLM at~\(t=0.3\):
    Schlieren type plots for divergence error (upper row) and contour
    plots for height (lower row); left with~\(c_\psi =
    2s_\text{max}\),~\(\delta = 2c_\psi\), right with~\(c_\psi =
    20s_\text{max}\),~\(\delta = 10^{-8}\).}
  \label{fig: pglmwin}
\end{figure}

Figure~\ref{fig: pglmwin} shows the results for~\(c_\psi =
2s_\text{max}\),~\(\delta = 2c_\psi\) and~\(c_\psi =
20s_\text{max}\),~\(\delta = 10^{-8}\). The thresholds for the
Schlieren type plots of the divergence errors are the same as in
Figure~\ref{fig:standardfail}. The divergence errors are concentrated
at the discontinuities, i.\,e.\ just where they are produced. Looking
at it in more detail, one finds that even outside the area confined by
the shock, some marbled structures are seen in the divergence
plots. This indicates, that the divergence errors are transported to
the outside. Although the divergence errors are higher for~\(c_\psi =
2s_\text{max}\),~\(\delta = 2c_\psi\), they are considerably better
than those without GLM (cf.\ Figure~\ref{fig:standardfail}). The
contour plots for the height are almost indistinguishable for both
sets of the parameters. Thus, we conclude that in general it is
sufficient to choose~\(c_\psi = 2s_\text{max}\),~\(\delta =
2c_\psi\). The slight loss in quality compared to~\(c_\psi =
20s_\text{max}\),~\(\delta = 10^{-8}\) is outweighed by the
considerable saving in computation time.

\begin{figure}
  \centering
  \input{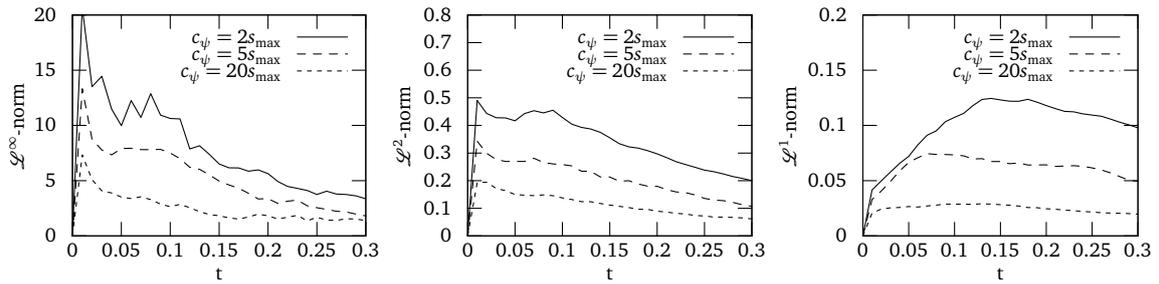}
  \caption{Generalized dam break with standard GLM: Comparison for
    different wave speeds; all cases computed with \(\delta =
    10^{-8}\).}
  \label{fig:glm-orig}
\end{figure}
In Figure~\ref{fig:glm-orig}, we present the error norms for different
choices of~\(c_\psi\) when used within the conservative (and, thus,
non-Galilean-invariant) GLM-system. Remember that in this case the
application of the Harten entropy fix on the artificial waves is
pointless since their wave speed is constant in space and determines
the time step. A comparison with Figure~\ref{fig:sepspeed} reveals
that the Galilean invariant (and, thus, non-conservative) GLM-system
is slightly better behaved, at least for this test case. With respect
to the higher computational effort for the eigensystem of the
conservative GLM-system, this is an encouraging result.

\begin{figure}
  \centering
  \input{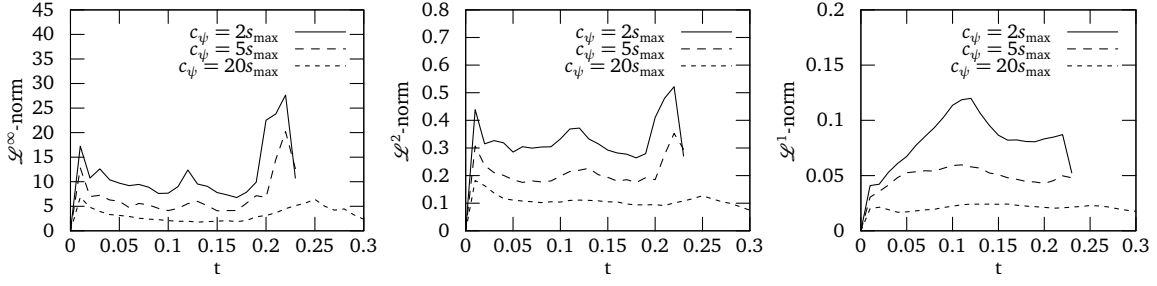}
  \caption{Generalized dam break: Non Galilean invariant, note the
    changed scaling for the \(\mathscr L^\infty\)-norm.}
  \label{fig:septrad}
\end{figure}

Finally, Figure~\ref{fig:septrad} shows that the system defined in
equation~\eqref{eq:15} is not able to provide any enhancement. Only
the standard equations lead to poorer results. 

\subsection{The de Sterck test}
\label{sec:de-sterck-test}

The De~Sterck test~\cite{hans-constrained} is a special configuration
for a shallow water MHD flow. It shows a strong tendency to develop
resonant phenomena and, thus, to single out numerical schemes which
are prone to divergence errors.  The test problem imposes a
supercritical, horizontal grid-aligned inflow on the left boundary of
a rectangular domain. The initial state in the lower half of the
domain, and also the boundary conditions at the left boundary, contain
a resonant mode. The initial data in the upper half are
\begin{equation}
  \label{eq-c:14}
  h=2\;, \qquad u=5.5\;, \qquad v=0\;, \qquad B_1 = 0.5\;, \qquad B_2 =0\;,
\end{equation}
and in the lower half
\begin{equation}
  \label{eq-c:15}
  h=1\;, \qquad u=4.5\;, \qquad v=0\;, \qquad B_1 = 2\;, \qquad B_2 =0\;.
\end{equation}
The gravitational constant is set to one. Since the discontinuity is
aligned with the grid, the initial data are discrete divergence free
for any reasonable difference operator.  We performed our tests on a
100\(\times\)100~grid for the domain~\([-1,1]\times[-1,1]\). We
compute up to time~\(t=4.8\), which is about six times the duration it
would take to arrive---physically---at a steady state.

\begin{figure}
  \centering
  \input{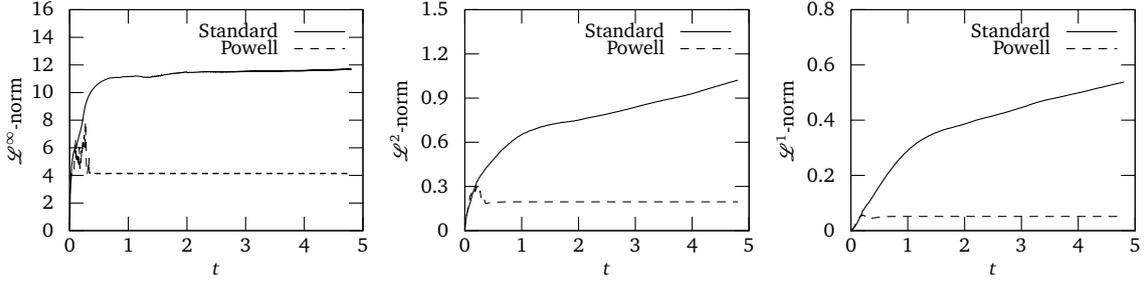}
  \caption{Error norms for de Sterck test without GLM.}
  \label{fig:hans-powell}
\end{figure}
\begin{figure}
  \centering
  \includegraphics[width=.7\linewidth]{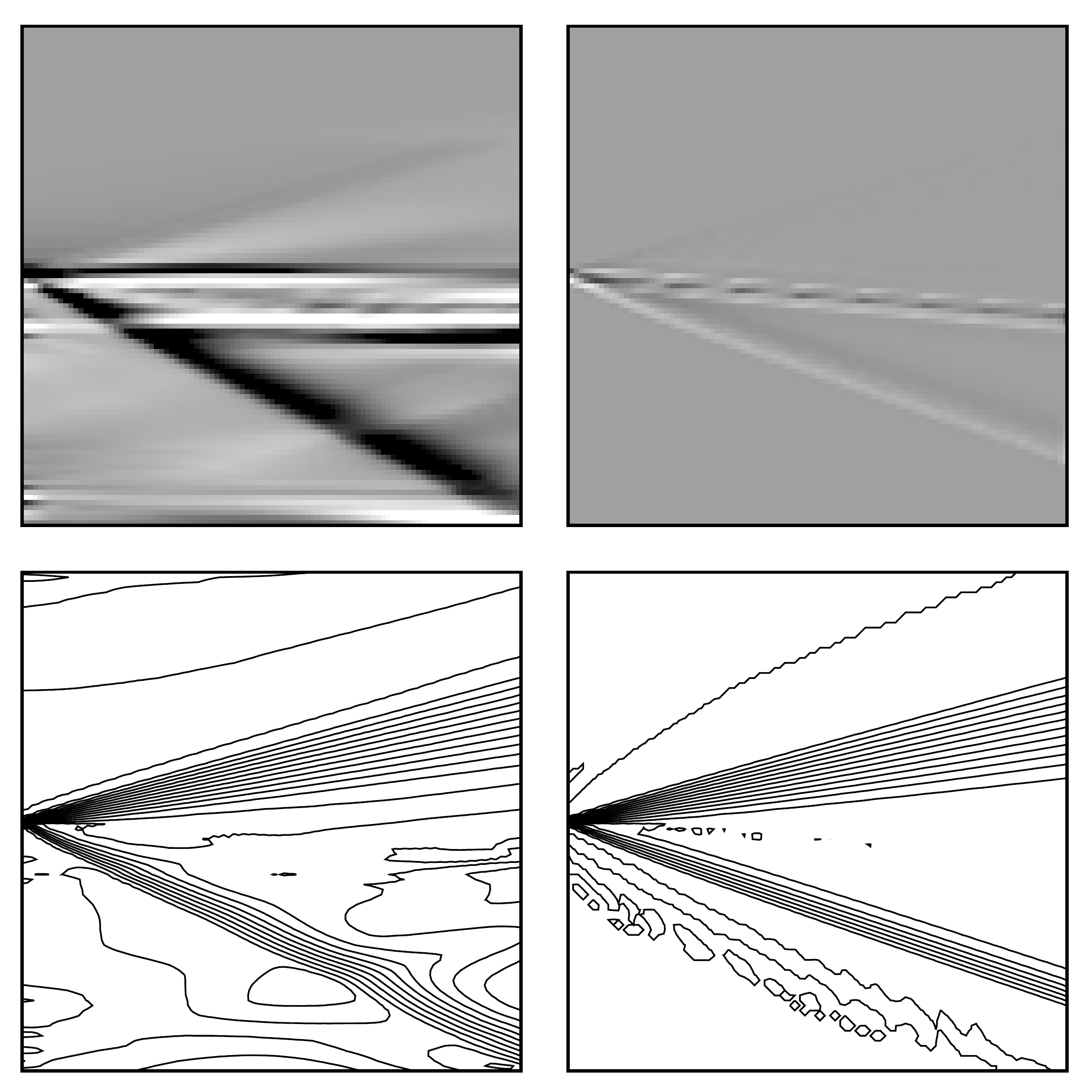}
  \caption{De Sterck test without GLM at~\(t=4.8\): Schlieren
    type plots for divergence error (upper row) and contour plots for
    height (lower row); left without divergence correction, right with
    Powell correction.}
  \label{fig:standard-hans}
\end{figure}
Figure~\ref{fig:hans-powell} shows the error norms for the standard
equations and for the Powell system. They indicate that for the Powell
system the divergence errors become stationary fairly soon, which in
turn means that the Powell system is suitable for steady state
computations. Without any divergence correction, the maximum of the
divergence errors is almost stationary. But as the other two norms
indicate, divergence errors of lower amplitude are produced during the
whole computation. The fact that the growth rate of the~\(\mathscr
L^2\)-norm is almost the same as for the~\(\mathscr L^1\)-norm
indicates that the newly generated divergence errors are still of
significant size.  The effect of the divergence errors on the solution
is illustrated in Figure~\ref{fig:standard-hans}. In this case, the
threshold for black and white in the divergence plots is set to~\(\pm
2\). While the physical solution is already stationary at~\(t=0.6\),
we continued the computation up
to~\(t=4.8\). Figure~\ref{fig:standard-hans} indicates that the Powell
system yields a good representation of the physical solution even
after such a long-term run, while for the standard system a bend in
one of the wave fronts is observed.

\begin{figure}
  \centering
  \input{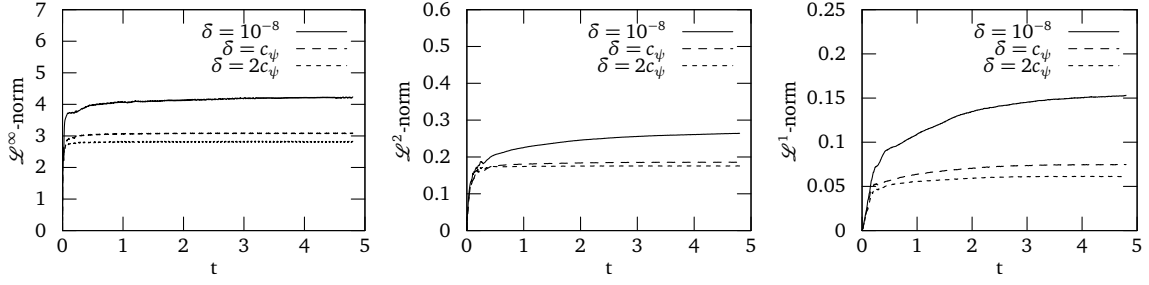}
  \caption{Error norms for de Sterck test with Powell+GLM and~\(c_\psi=2s_\text{max}\) for different viscosities.}
  \label{fig:hans-visc}
\end{figure}
\begin{figure}
  \centering
  \input{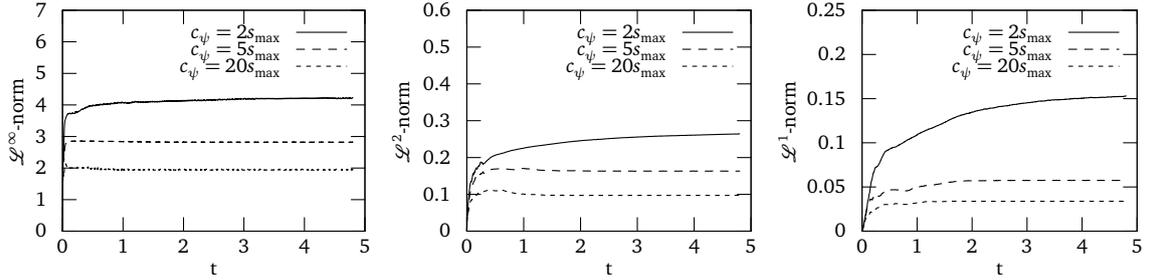}
  \caption{Error norms for de Sterck test with Powell+GLM and
    different choices for the artificial wave speed.}
  \label{fig:hans-pglm}
\end{figure}
\begin{figure}
  \centering
  \includegraphics[width=.7\linewidth]{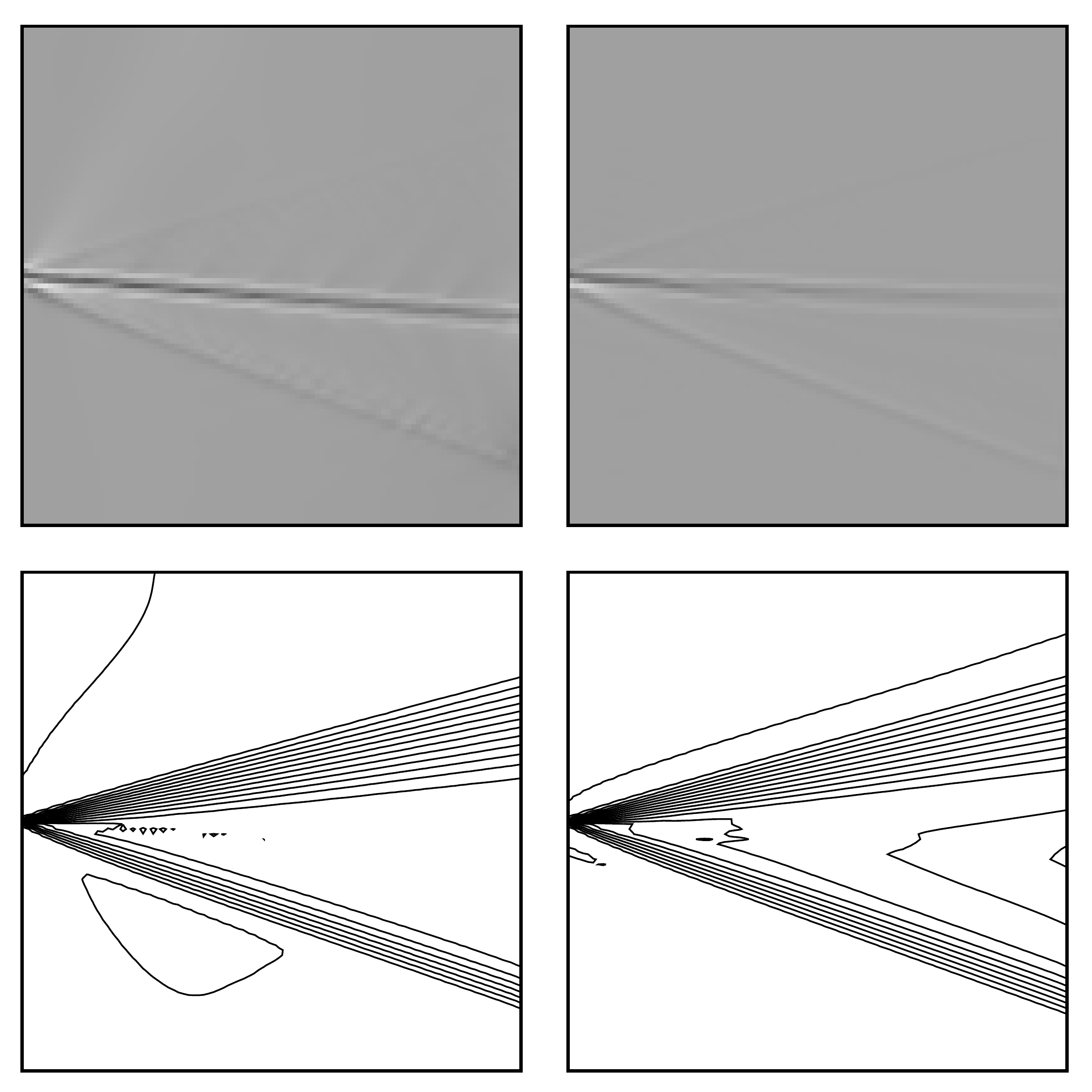}
  \caption{De Sterck test with Powell+GLM at~\(t=4.8\): Schlieren
    type plots for divergence error (upper row) and contour plots for
    height (lower row); left without divergence correction, right with
    Powell correction.}
  \label{fig:pglm-hans}
\end{figure}

In Figures~\ref{fig:hans-visc} and~\ref{fig:hans-pglm}, we show
comparisons for different settings of viscosity and speed of the
artificial waves in the Powell-GLM system. Like for the generalized
dam break, the choice~\(c_\psi = 2s_\text{max}\),~\(\delta = 2c_\psi\)
results in about the same error norms as~\(c_\psi =
s_\text{max}\),~\(\delta = 10^{-8}\). Increasing the wave speed
to~\(c_\psi = 20s_\text{max}\) only reduces the errors by a factor
less than~\(3/2\). In Figure~\ref{fig:pglm-hans}, we show the results
for the height. The results with~\(c_\psi = 2s_\text{max}\),~\(\delta
= 2c_\psi\) are slightly better than for the pure Powell correction
(Figure~\ref{fig:standard-hans}). The gain in quality for~\(c_\psi =
20s_\text{max}\),~\(\delta = 10^{-8}\) is however rather small.

\begin{figure}
  \centering
  \input{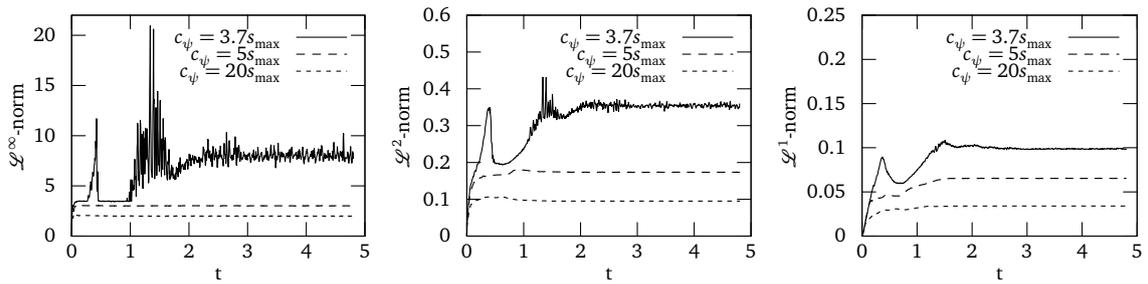}
  \caption{Error norms for de Sterck test with pure GLM and different
    choices for the artificial wave speed. Note the different scaling
    for the~\(\mathscr L^\infty\)-norm.}
  \label{fig:hans-glm}
\end{figure}
For the pure GLM correction, the results are worse than for the
combined Powell-GLM system. We had to choose~\(c_\psi\geq 3.7\) to
arrive at a stable scheme. For higher values of~\(c_\psi\), the errors
are still slightly higher than when using Powell and GLM together.
Note that all presented results are still considerably improved in
comparison to what one could achieve without any divergence
correction~\cite{habil-kemm,orig-div}. For the system that combines
the eigenvalues of standard GLM and the eigenvectors of Powell GLM
(cf.\ equations~\eqref{eq:15} and~\eqref{eq:30}), even with~\(c_\psi
=20 s_\text{max}\), the computation failed due to excessive divergence
errors.

\section{Conclusions}
\label{sec:conclusions}

In this paper, we discussed several forms of the shallow water
magnetohydrodynamics equations and their Roe-type discretizations. Two
of these models, the standard form and the equations with the standard
application of the hyperbolic Generalized Lagrange Multiplier (GLM)
correction, are conservation systems. We also discussed two Galilean
invariant forms: the Powell system and the Powell system with
hyperbolic GLM. A special case is the modification described at the
end of Section~\ref{sec:equations-with-glm}. It combines the
eigenvalues of standard GLM with the eigenvectors of the Galilean
invariant GLM system, a strategy which reduces the computational
effort for the eigensystem when compared to standard
GLM. Unfortunately, this version yields poor numerical results and is,
therefore, not suitable for practical application.

As found in Section~\ref{sec:govern-equat-comb}, in the absence of
magnetic monopoles, the waves are essentially the same for the
conservative ant the non-conservative models. (Remember that all
artificial waves vanish in that case.) Thus, the best one can do is to
apply a solver based on the characteristic decomposition. Thus, we
resorted to Roe-type solvers. Furthermore, the solver for any of the
non-conservative models should also coincide with a conservative
solver for a solenoidal magnetic field.  We showed how the usual way
to obtain Roe matrices for the non-conservative systems not only
involves computationally expensive logarithms of the physical
quantities, but also leads to fill-in in the matrix structure, which
in turn results in a highly complicated and computationally even more
expensive eigensystem.  Without magnetic monopoles, or even artificial
monopoles resulting from numerical errors, all physics is represented
by the standard equations~\eqref{eq-smhd:65}\,--\,\eqref{eq:2}. We
therefore concluded that it is reasonable to resort to the system
matrix evaluated in the Roe mean value for the standard equations. The
requirement for the solver to coincide with a conservative solver for
a solenoidal magnetic field is still satisfied.

Our numerical tests in the last section confirm the superiority of the
non-conservative models over the conservative models, especially the
standard model. Even the pure the Powell system without additional GLM
performs much better than the standard system. As was seen in the case
of the generalized dam break problem, this is mainly due to its full
hyperbolicity and thus the prevention of any resonant behaviour. In
the highly advection dominated De~Sterck test, it even performed
similar to some versions of the combined Powell+GLM system. From the
results in Section~\ref{sec:numerical-tests}, we conclude that in such
cases, i.\,e.\ when the advection allows an efficient transport of the
divergence errors out of the computational domain, it is sufficient to
employ the Powell system. In terms of computational cost, it is even
preferable, since it allows for longer time steps than the GLM
versions. But in general, it is favourable to use the combined
Powell+GLM system with~\(c_\psi = 2s_\text{max}\),~\(\delta =
2c_\psi\), which is sufficient to provide high quality, physically
relevant results.

\minisec{Acknowledgments}
This work was supported by DFG research grant KE\,1420/3-1.

\bibliographystyle{amsplain}
\bibliography{smhd-noncons}

\providecommand{\bysame}{\leavevmode\hbox to3em{\hrulefill}\thinspace}
\providecommand{\MR}{\relax\ifhmode\unskip\space\fi MR }
\providecommand{\MRhref}[2]{%
  \href{http://www.ams.org/mathscinet-getitem?mr=#1}{#2}
}
\providecommand{\href}[2]{#2}
\begin{thebibliography}{10}

\bibitem{balsdiv}
Dinshaw~S.\ Balsara, \emph{Divergence-free adaptive mesh refinement for
  magnetohydrodynamics}, J.\ Comput.\ Phys. \textbf{174} (2001), 614--648.

\bibitem{spicer}
Dinshaw~S.\ Balsara and Daniel~S.\ Spicer, \emph{A staggered mesh algorithm
  using high order {G}odunov fluxes to ensure solenoidal magnetic fields in
  magnetohydrodynamic simulations}, J.\ Comput.\ Phys. \textbf{149} (1999),
  270--292.

\bibitem{besse05}
Nicolas Besse and Dietmar Kr\"oner, \emph{Convergence of locally
  divergence-free discontinuous-{Galerkin} methods for the induction equations
  of the {2D-MHD} system}, M2AN Math.\ Model.\ Numer.\ Anal. \textbf{39}
  (2005), no.~6, 1177--1202.

\bibitem{brackbill1}
Jeremiah~U. Brackbill and D.~C. Barnes, \emph{The effect of nonzero $\nabla
  \cdot {B}$ on the numerical solution of the magnetohydrodynamic equations},
  J.\ Comput.\ Phys. \textbf{35} (1980), 426--430.

\bibitem{clawpack}
\emph{Clawpack (conservation laws package)},
  http://www.amath.washington.edu/~claw.

\bibitem{Crockett}
Robert~K. Crockett, Phillip Colella, Robert~T. Fisher, Richard~I. Klein, and
  Christopher~F. McKee, \emph{An unsplit, cell-centered godunov method for
  ideal {MHD}}, J.\ Comput.\ Phys. \textbf{203} (2005), no.~2, 422--448.

\bibitem{dafermospaper}
Constantine~M.\ Dafermos, \emph{Quasilinear hyperbolic systems with
  involutions}, Arch.\ Ration.\ Mech.\ Anal. \textbf{94} (1986), 373--389.

\bibitem{dafermosbuch}
\bysame, \emph{Hyperbolic conservation laws in continuum physics}, Springer,
  Berlin, Heidelberg, 2000.

\bibitem{divpaper}
Andreas Dedner, Friedemann Kemm, Dietmar Kr\"oner, Claus-Dieter Munz, Thomas
  Schnitzer, and Matthias Wesenberg, \emph{Hyperbolic divergence cleaning for
  the {MHD} equations}, J.\ Comput.\ Phys. \textbf{175} (2002), no.~2,
  645--673.

\bibitem{hllem}
Bernd Einfeldt, \emph{{On {G}odunov-type methods for gas dynamics.}}, SIAM J.
  Numer. Anal. \textbf{25} (1988), no.~2, 294--318 (English).

\bibitem{hawley88}
C.~R. Evans and John~F. Hawley, \emph{Simulation of general relativistic
  magnetohydrodynamic flows: A constrained transport method}, Astrophys.\ J.
  \textbf{332} (1988), 659.

\bibitem{fuchs-split}
F.G. Fuchs, S.~Mishra, and N.H. Risebro, \emph{{Splitting based finite volume
  schemes for ideal {MHD} equations.}}, J.\ Comput.\ Phys. \textbf{228} (2009),
  no.~3, 641--660 (English).

\bibitem{fuchs-induction}
Franz~G. Fuchs, Kenneth~H. Karlsen, Siddharta Mishra, and Nils~H. Risebro,
  \emph{{Stable upwind schemes for the magnetic induction equation.}}, ESAIM,
  Math. Model. Numer. Anal. \textbf{43} (2009), no.~5, 825--852 (English).

\bibitem{gilman-smhd}
Peter~A. Gilman, \emph{Magnetohydrodynamic ``shallow water'' equations for the
  solar tachocline}, Astrophys J.\ Letters \textbf{544} (2000), no.~2, L79.

\bibitem{godunov1972symmetric}
S.K. Godunov, \emph{Symmetric form of the magnetohydrodynamic equation},
  Chislennye Metody Mekh. Sploshnoi Sredy \textbf{3} (1972), no.~1, 26--34.

\bibitem{harten-tvd}
Amiram Harten, \emph{High resolution schemes for hyperbolic conservation laws},
  J.\ Comput.\ Phys. \textbf{49} (1983), no.~3, 357--393.

\bibitem{harten-hyman}
Amiram Harten and James~M. Hyman, \emph{Self adjusting grid methods for
  one-dimensional hyperbolic conservation laws}, J.\ Comput.\ Phys. \textbf{50}
  (1983), no.~2, 235--269.

\bibitem{limiter}
Friedemann Kemm, \emph{A comparative study of {TVD} limiters\,--\,well known
  limiters and an introduction of new ones}, Internat.\ J.\ Numer.\ Methods
  Fluids \textbf{67} (2011), no.~4, 404--440.

\bibitem{kemm-santiago}
\bysame, \emph{{CFL}-number-dependent {TVD}-limiters}, Numerical Methods for
  Hyperbolic Equations: Theory and Applications (Elena Vázquez-Cendón, Arturo
  Hidalgo, Pilar García-Navarro, and Luis Cea, eds.), CRC Press, 2012,
  Proceedings of the international conference on Numerical Methods for
  Hyperbolic Equations: Theory and Applications, Santiago de Compostela, Spain,
  4--9 July 2011, pp.~277--283.

\bibitem{orig-div}
\bysame, \emph{On the origin of divergence errors in mhd simulations and
  consequences for numerical schemes}, Commun.\ Appl.\ Math.\ Comput.\ Sci.
  \textbf{8} (2013), no.~1, 1--38.

\bibitem{habil-kemm}
\bysame, \emph{Contributions to the numerical simulation of gas, shallow water,
  and plasma flows -- wave-wise treatment of order and viscosity},
  Habilitationsschrift, Brandenburgische Technische Universität, 2014.

\bibitem{porquerolles}
Friedemann Kemm, Yong-Joong Lee, Claus-Dieter Munz, and Rudolf Schneider,
  \emph{{Divergence cleaning in finite-volume computations for electromagnetic
  wave propagations.}}, {Finite volumes for complex applications III. Problems
  and perspectives. Papers from the 3rd symposium of finite volumes for complex
  applications, Porquerolles, France, June 24-28} (Rapha\'ele Herbin and
  Dietmar Kr{\"o}ner, eds.), Hermes Penton Science, London, 2002, pp.~575--582
  (English).

\bibitem{pasadena}
Y.J. Lee, F.~Kemm, C.-D. Munz, and R.~Schneider, \emph{Physical symmetries and
  hyperbolic {GLM} divergence correction scheme for {Maxwell} and mhd
  equations}, Hyperbolic Problems: Theory, Numerics, Applications (ThomasY. Hou
  and Eitan Tadmor, eds.), Springer Berlin Heidelberg, 2003, pp.~685--694
  (English).

\bibitem{anume}
Y.J. Lee, R.~Schneider, C.-D. Munz, and F.~Kemm, \emph{Hyperbolic {GLM} scheme
  for elliptic constraints in computational electromagnetics and {MHD}},
  Analysis and Numerics for Conservation Laws (Gerald Warnecke, ed.), Springer
  Berlin Heidelberg, 2005, pp.~385--404 (English).

\bibitem{Marder87}
Barry Marder, \emph{A method incorporating gau\ss' law into electromagnetic
  {PIC} codes}, J.\ Comput.\ Phys. \textbf{68} (1987), 48--55.

\bibitem{mishra-I}
Siddhartha Mishra and Eitan Tadmor, \emph{Constraint preserving schemes using
  potential-based fluxes. {I}. multidimensional transport equations}, Commun.\
  Comput.\ Phys \textbf{9} (2010), 688--710.

\bibitem{mishra-II}
Siddhartha {Mishra} and Eitan {Tadmor}, \emph{{Constraint preserving schemes
  using potential-based fluxes. {II}: Genuinely multidimensional systems of
  conservation laws.}}, {SIAM J.\ Numer.\ Anal.} \textbf{49} (2011), no.~3,
  1023--1045 (English).

\bibitem{mishra-III}
\bysame, \emph{{Constraint preserving schemes using potential-based fluxes.
  {III}: Genuinely multi-dimensional schemes for {MHD} equations.}}, {ESAIM,
  Math.\ Model.\ Numer.\ Anal.} \textbf{46} (2012), no.~3, 661--680 (English).

\bibitem{maxjcp99}
Claus-Dieter Munz, Pascal Omnes, Rudolf Schneider, Eric Sonnendr\"ucker, and
  Ursula Voss, \emph{{D}ivergence correction techniques for {M}axwell solvers
  based on a hyperbolic model}, J.\ Comput.\ Phys. \textbf{161} (1999),
  484--511.

\bibitem{timedomain}
Claus-Dieter Munz, Rodolf Schneider, and Ursula Vo\ss{}, \emph{A finite volume
  method for the {M}axwell equations in the time domain}, SIAM J.\ Sci.\
  Comput. \textbf{22} (2000), no.~2, 449--475.

\bibitem{eric}
Claus-Dieter Munz, Rudolf Schneider, Eric Sonnendr\"ucker, and Ursula Voss,
  \emph{{M}axwell's equations when the charge conservation is not
  {satisf\/ied}}, C.\ R.\ Acad.\ Sci.\ Paris \textbf{328} (1999), no.~S'erie I,
  431--436.

\bibitem{powell95}
Kenneth~G. Powell, Phil~L. Roe, R.~\~S.\ Myong, T.~Gombosi, and D.~de~Zeeuw,
  \emph{An upwind scheme for magnetohydrodynamics}, Workshop Methodes
  numeriques pour la M.H.D. (Luminy, France), 1995.

\bibitem{linde99}
Kenneth~G. Powell, Philipp~L. Roe, Timur Linde, Tamas~I. Gombosi, and Darren~L.
  DeZeeuw, \emph{A solution-adaptive upwind scheme for ideal
  magnetohydrodynamics}, J.\ Comput.\ Phys. \textbf{154} (1999), 284--309.

\bibitem{Qamar2006132}
Shamsul {Qamar} and Gerald {Warnecke}, \emph{{Application of space-time {CE/SE}
  method to shallow water magnetohydrodynamic equations.}}, {J.\ Comput.\
  Appl.\ Math.} \textbf{196} (2006), no.~1, 132--149 (English).

\bibitem{roe-orig}
Philip~L.\ Roe, \emph{Approximate {R}iemann solvers, parameter vectors, and
  difference schemes}, J.\ Comput.\ Phys. \textbf{43} (1981), no.~2, 357--372.

\bibitem{rossmanith}
James~A.\ Rossmanith, \emph{{A wave propagation method with constrained
  transport for ideal and shallow water magnetohydrodynamics}}, Ph.D. thesis,
  University of Washington, 2002.

\bibitem{rossmanith06}
James~A.\ Rossmanith, \emph{An unstaggered high-resolution constrained
  transport method for magnetohydrodynamic flows}, SIAM J.\ Sci.\ Comput.
  \textbf{28} (2006), no.~5, 1766--1797.

\bibitem{hans1}
Hans~De Sterck, \emph{Hyperbolic theory of the ``shallow water''
  magnetohydrodynamics equations}, Phys.\ Plasmas \textbf{8} (2001), no.~7,
  3293--3304.

\bibitem{hans-constrained}
\bysame, \emph{Multi-dimensional upwind constrained transport on unstructured
  grids for ``shallow water'' magnetohydrodynamics}, AIAA 2001-2623, 2001.

\bibitem{Manuel}
Manuel Torrilhon, \emph{Locally divergence-preserving upwind finite volume
  schemes for magnetohydrodynamic equations}, SIAM J.\ Sci.\ Comput.
  \textbf{26} (2005), no.~4, 1166--1191.

\bibitem{manfey}
Manuel Torrilhon and Michael Fey, \emph{Constraint-preserving upwind methods
  for multidimensional advection equations}, SIAM J.\ Numer.\ Anal. \textbf{42}
  (2004), no.~4, 1694--1728.

\bibitem{Toumi1992360}
I.\ Toumi, \emph{A weak formulation of {Roe's} approximate {Riemann} solver},
  Journal of Computational Physics \textbf{102} (1992), no.~2, 360 -- 373.

\bibitem{waagan}
Knut Waagan, \emph{A positive {MUSCL-Hancock} scheme for ideal
  magnetohydrodynamics}, J.\ Comput.\ Phys. \textbf{228} (2009), no.~23,
  8609--8626.

\end{thebibliography}

\end{document}